\documentclass[
    prd,
    aps,
    amsfonts,
    onecolumn,
    groupedaddress,
    nofootinbib,
    floatfix,
    raggedbottom,
    11pt
]{revtex4-2}

\usepackage[utf8]{inputenc}
\usepackage[T1]{fontenc}
\usepackage{microtype}

\usepackage{amsmath}
\usepackage{amssymb}
\usepackage{mathrsfs}
\usepackage{nicefrac}
\usepackage{dsfont}
\usepackage{bm}
\usepackage{slashed}

\usepackage{graphicx}
\graphicspath{{./Images/}}

\usepackage{tikz}
\usepackage[compat=1.1.0]{tikz-feynman}
\pgfkeys{/tikzfeynman/warn luatex=false}

\usepackage[font=small,labelfont=bf]{caption}
\usepackage{subcaption}

\usepackage{booktabs}
\usepackage{array}
\usepackage{dcolumn}
\usepackage{tabularx}
\usepackage{ragged2e}

\usepackage{float}
\usepackage{placeins}

\usepackage{xcolor}
\usepackage{hyperref}

\hypersetup{
    colorlinks=true,
    linkcolor=blue,
    citecolor=blue,
    urlcolor=blue
}

\begin{document}

\title{Mixed Freeze-In and Freeze-Out Histories and Dark-Sector Decays in a
\texorpdfstring{$\mathbb{Z}_4$}{Z4} Two-Scalar Model}

\author{J. P. Carvalho-Corrêa}
\email{jpcarv-15897@ufmg.br}

\author{B. A. Couto e Silva}
\email{brunoaces@ufmg.br}

\author{B. L. S\'anchez-Vega}
\email{bruce@fisica.ufmg.br}

\affiliation{Departamento de F\'isica, UFMG, Belo Horizonte, MG 31270-901, Brazil.}

\begin{abstract}
We study dark matter production in a renormalisable
\(\mathbb{Z}_4\) Higgs-portal model containing a complex scalar \(S_A\)
and a real scalar \(S_B\), focusing on scenarios in which one or both
states remain out of thermal equilibrium with the Standard Model bath.
We follow their coupled evolution, including conversion,
semi-annihilation, and the decay \(S_B\to S_A S_A\) when kinematically
allowed. We impose theoretical consistency conditions, limits on
invisible Higgs decays, and LZ direct-detection bounds, taking into
account the relic fraction of each component. We also check that the
WIMP/FIMP assignments agree with the interaction rates. When both
states are stable, we find viable FIMP--FIMP and mixed WIMP--FIMP
scenarios. In the mixed cases, the WIMP-like state remains constrained
by its Higgs-portal interactions, with the surviving points mainly
concentrated near the Higgs resonance. The FIMP-like state can,
however, provide a substantial fraction of the relic density with
a negligible direct-detection signal. For \(M_{S_B}>2M_{S_A}\),
the decay of the heavier scalar leads to three different production
histories: SuperWIMP production, injection-assisted freeze-out, and
sequential freeze-in. Which mechanism occurs depends on how the
parent and daughter particles were produced. These decays can
increase the abundance of a WIMP after freeze-out or contribute
to the final abundance of a FIMP, changing the relation between
the daughter's Higgs coupling and its relic density. Since the
primary decays produce no Standard Model particles, the usual
electromagnetic and hadronic BBN bounds on visible decay products
do not apply directly. We also find \(r_{\rm BBN}\ll1\) throughout
the retained sample. To assess possible effects on structure
formation, we use the fraction of dark matter produced through
decays and an equivalent thermal-WDM mass as approximate diagnostics.
Some configurations with late decays are flagged as potentially
warm, whereas the sequential freeze-in sample satisfies
\(m_{\rm WDM}^{\rm eq}\gtrsim1~\mathrm{MeV}\).
\end{abstract}

\maketitle

\newpage
\tableofcontents

\newpage

\section{Introduction}
\label{sec:introduction}

A wide range of cosmological and astrophysical observations firmly
establishes the existence of non-baryonic dark matter (DM), which
dominates the matter budget of the Universe and plays a central
role in structure formation~\cite{Planck2018,RubinFord1970,
SofueRubin2001,Clowe2006}. Despite this robust gravitational evidence,
the microscopic nature of DM remains unknown~\cite{PDGDM,Bertone2005}.
Since the Standard Model (SM) offers no viable candidate, the DM
puzzle remains one of the sharpest motivations for physics beyond
the SM.

Theoretical benchmarks and experimental search strategies have
traditionally focused on scenarios in which the dark sector is
effectively described by a single stable particle~\cite{
Jungman1996,Bertone2005,Arcadi2018,Sanchez-Vega:2014rka,
Sanchez-Vega:2015qva,Dias:2025oyb}. Single-component scenarios
already encompass a broad range of candidates and production
histories, including non-thermal realizations such as axion dark
matter in extended gauge sectors~\cite{Montero:2017yja}.
Multi-component scenarios broaden this landscape further and arise
naturally in frameworks where stability is enforced by symmetries
larger than $\mathbb{Z}_2$~\cite{Zurek2009,Aoki2012,Capucha,
Pukhov,Profumo2009,EschKlasenYaguna2014,Dienes2012,DiazSaez_2021}.
Such models also allow processes such as semi-annihilation and
conversion between dark states, which can change the relic abundance
of each component and affect the resulting experimental
constraints~\cite{GriestSeckel1991,DEramoThaler2010,
Belanger:2012zr,ThomasHambye_2009}. In scalar models based on
$\mathbb{Z}_{2n}$ symmetries, the charge assignments determine
which interactions are allowed and, together with the mass
hierarchy, which particles remain stable~\cite{
YagunaZapata2020,Yaguna2021}.

In Ref.~\cite{CarvalhoCorrea:2025Z2n}, we studied two-component
scalar dark matter in the $\mathbb{Z}_4$, $\mathbb{Z}_6(13)$, and
$\mathbb{Z}_6(23)$ models, assuming that both components were
thermal WIMPs. That analysis combined updated direct-detection
limits with theoretical constraints from boundedness from below,
perturbative unitarity, and perturbativity, including one-loop
effects. We found that the LZ spin-independent limits~\cite{
LZ:2025PRL_4p2ty} exclude large regions of parameter space,
even after accounting for the relic fraction of each component.
In particular, some parameter points satisfy the rescaled
limit for each particle separately but fail the combined
constraint when both contributions are considered. This leads
to the question: \emph{How do the cosmological evolution and
phenomenology of the minimal $\mathbb{Z}_4$ model change if one
or both dark states are produced without reaching thermal
equilibrium?}

To address this question, we study the renormalisable
$\mathbb{Z}_4$ model with a complex scalar $S_A$ and a real
scalar $S_B$, allowing one or both states to be produced
through freeze-in. We consider both the case in which the two
scalars are stable and the mass hierarchy $M_{S_B}>2M_{S_A}$,
where $S_B$ can decay into pairs of $S_A$ particles or
antiparticles. We solve the coupled Boltzmann equations,
including Higgs-portal production and interactions between
the dark scalars. We examine how the relic abundance is
shared when both states are stable and which regions of
parameter space remain consistent with theoretical and
experimental constraints. When $S_B$ decays, we study its
contribution to the final $S_A$ abundance and assess possible
effects on the expansion rate and structure formation.

The paper is organised as follows. Section~\ref{sec:model}
introduces the model and its interactions. Section~\ref{sec:cosmo}
presents the coupled Boltzmann equations and the experimental
and cosmological constraints. Section~\ref{sec:results}
describes the numerical scan and discusses the stable and
decay-mediated scenarios, including the check of the WIMP/FIMP
assignments. We conclude in Sec.~\ref{sec:conclusions}.

\section{The Model}
\label{sec:model}

We consider a renormalisable scalar extension of the SM invariant under a discrete $\mathbb{Z}_4$ symmetry, introduced
in Ref.~\cite{Yaguna2021} as a minimal setup for two-component scalar dark matter. All SM fields are taken to be neutral
under $\mathbb{Z}_4$, while the particle content is augmented by a complex scalar $S_A$ and a real scalar $S_B$.
Denoting the $\mathbb{Z}_4$ generator by $\omega_4=e^{i\pi/2}$, the new fields transform as
\begin{equation}
S_A \;\to\; \omega_4\, S_A= i\,S_A,
\qquad
S_B \;\to\; \omega_4^2\, S_B = -S_B,
\end{equation}
i.e.\ $S_A$ carries charge $1$ and $S_B$ charge $2$ modulo $4$. These charge assignments render $S_A$ stable,
whereas the stability of $S_B$ depends on the mass hierarchy. For $M_{S_B}<2M_{S_A}$, both fields are stable and
constitute a two-component dark sector. For $M_{S_B}>2M_{S_A}$, the cubic interaction $S_A^2S_B$ allowed by the
symmetry opens the decay channels $S_B\to S_A S_A$ and $S_B\to S_A^\dagger S_A^\dagger$, so that $S_B$ becomes
unstable while $S_A$ remains the final stable relic. This distinction defines the stable and decay-mediated regimes
analysed below.

The most general renormalisable Lagrangian consistent with the SM gauge symmetry and these charge assignments is
\begin{equation}
\label{eq:lag}
\mathcal{L}
= \mathcal{L}_{\text{SM}}
+ (\partial_\mu S_A^*)(\partial^\mu S_A)
+ \frac{1}{2} (\partial_\mu S_B)(\partial^\mu S_B)
- V(H,S_A,S_B),
\end{equation}
with scalar potential
\begin{equation}
\label{eq:pot}
\begin{split}
V(H,S_A,S_B) =\;&
-\mu_H^2 |H|^2 + \lambda_H |H|^4
-\mu_A^2 |S_A|^2 + \lambda_A |S_A|^4
-\frac{1}{2}\mu_B^2 S_B^2 + \lambda_B S_B^4 \\
&+ \lambda_{HA} |H|^2 |S_A|^2
+ \frac{1}{2}\lambda_{HB} |H|^2 S_B^2
+ \lambda_{AB} |S_A|^2 S_B^2 \\
&+ \frac{1}{2}\Big(\mu_{S1}\, S_A^2 S_B + \lambda_{S4}\, S_A^4\Big) + \text{h.c.}\, .
\end{split}
\end{equation}

Throughout this work we assume CP conservation in the scalar sector. In particular, we choose a basis where
$\mu_{S1}$ and $\lambda_{S4}$ are real, so that the last line of Eq.~\eqref{eq:pot} is equivalently the real
combination $\frac{\mu_{S1}}{2}(S_A^2+(S_A^\dagger)^2)S_B+\frac{\lambda_{S4}}{2}(S_A^4+(S_A^\dagger)^4)$.

For later use in the coupled Boltzmann system, it is convenient to identify the interactions associated with the
leading number-changing processes. The Higgs-portal couplings $\lambda_{HA}$ and $\lambda_{HB}$ connect $S_A$ and
$S_B$ to the SM through Higgs-mediated interactions. The coupling $\lambda_{AB}$ provides the quartic contact
interaction relevant for the conversion process $S_A S_A^\dagger \leftrightarrow S_B S_B$, while the trilinear
coupling $\mu_{S1}$ enables semi-annihilation and, when kinematically allowed, the decay
$S_B\to S_A S_A$ and its conjugate channel. The corresponding reaction rates are discussed in
Sec.~\ref{sec:cosmo}.

After electroweak symmetry breaking (EWSB), the Higgs acquires a vacuum expectation value
$\langle H \rangle=(0,v)^{T}/\sqrt{2}$. We focus on a $\mathbb{Z}_4$-preserving vacuum with
$\langle S_A \rangle=\langle S_B \rangle=0$, which prevents mixing between the Higgs and the dark scalars. The physical
masses are $M_h^2=2\lambda_H v^2$ and
\begin{equation}
M_{S_A}^2 = \frac{1}{2}\lambda_{HA} v^2 - \mu_A^2,
\qquad
M_{S_B}^2 = \frac{1}{2}\lambda_{HB} v^2 - \mu_B^2.
\end{equation}

In practice, we trade the quadratic parameters $\mu_A^2$ and $\mu_B^2$ for the physical masses. A convenient set of
independent Lagrangian parameters is then given by the two dark masses $(M_{S_A},M_{S_B})$, the trilinear coupling
$\mu_{S1}$, and the six dimensionless couplings
$(\lambda_A,\lambda_B,\lambda_{HA},\lambda_{HB},\lambda_{AB},\lambda_{S4})$. The subset of parameters varied in the
numerical scans, together with the couplings held fixed, is specified in Sec.~\ref{sec:results}.

Theoretical consistency requires the scalar potential to be bounded from below (BFB), the dimensionless scalar
couplings to remain perturbative, and the high-energy $2\to2$ scalar scattering amplitudes to satisfy perturbative
unitarity. Following Refs.~\cite{Kannike2012,CarvalhoCorrea:2025Z2n}, the BFB conditions along the individual field
directions are
\begin{equation}
\label{eq:estabilidade_diagonal}
\lambda_H > 0,
\qquad
\lambda_A - |\lambda_{S4}| > 0,
\qquad
\lambda_B > 0,
\end{equation}
while the pairwise copositivity conditions are
\begin{align}
\overline{\lambda}_{AB}
&\equiv
\lambda_{AB}
+2\sqrt{(\lambda_A-|\lambda_{S4}|)\lambda_B}
>0,
\label{eq:condAB}
\\
\overline{\lambda}_{HA}
&\equiv
\lambda_{HA}
+2\sqrt{\lambda_H(\lambda_A-|\lambda_{S4}|)}
>0,
\label{eq:condHA}
\\
\overline{\lambda}_{HB}
&\equiv
\lambda_{HB}
+4\sqrt{\lambda_H\lambda_B}
>0.
\label{eq:condHB}
\end{align}
These conditions must be supplemented by an additional copositivity condition involving all three field
directions, which is given in Appendix~\ref{app:unitarity}. We impose the complete set of BFB conditions at tree
level.

Perturbativity is imposed through $|\lambda_i|\leq 4\pi$ for all dimensionless scalar couplings. Perturbative
unitarity is enforced through the standard partial-wave condition \(|\Re(a_0)|\leq 1/2\) for the
high-energy $2\to2$ scalar scattering amplitudes. The explicit definitions and normalisation of the
scattering-matrix eigenvalues, together with their corresponding bounds, are given in
Appendix~\ref{app:unitarity}. The appendix also presents the cubic equation whose roots determine the remaining
eigenvalue combinations. All the theoretical constraints described above are enforced point-by-point in our
numerical analysis.

\section{Cosmological Evolution and Phenomenology}
\label{sec:cosmo}

\begin{figure}[!t]
    \centering
    \captionsetup[subfigure]{font=small,skip=2pt}
    \captionsetup{font=small}

    \begin{subfigure}{0.75\textwidth}
        \centering
        \includegraphics[width=\linewidth]{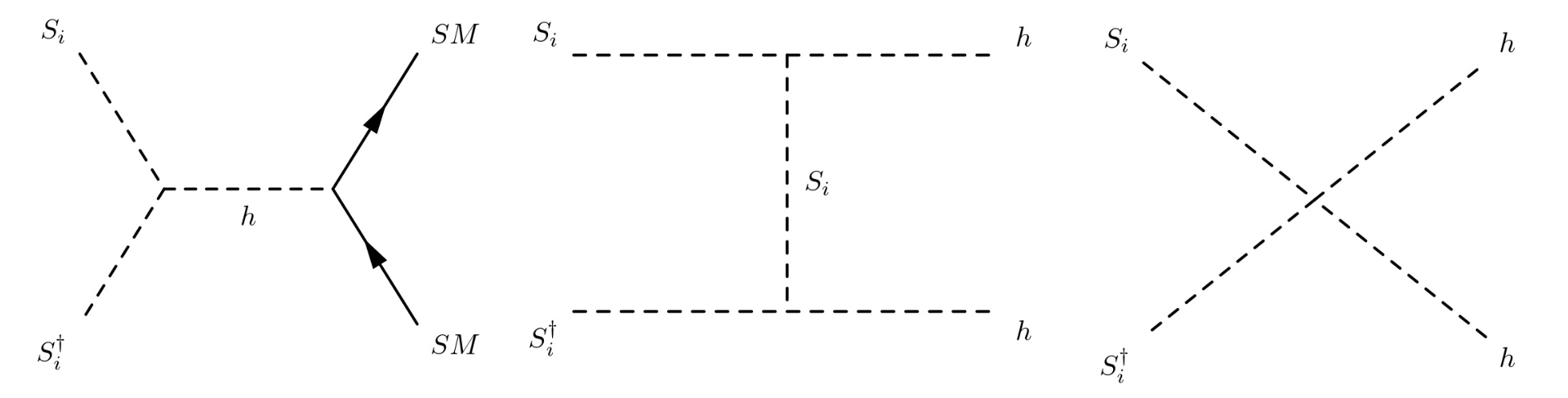}
        \caption{\justifying Higgs-portal annihilation of \(S_i\) (\(i=A,B\)) into SM final states.}
        \label{fig:deteccdire}
    \end{subfigure}

    \vspace{0.25em}

    \begin{subfigure}{0.6\textwidth}
        \centering
        \includegraphics[width=\linewidth]{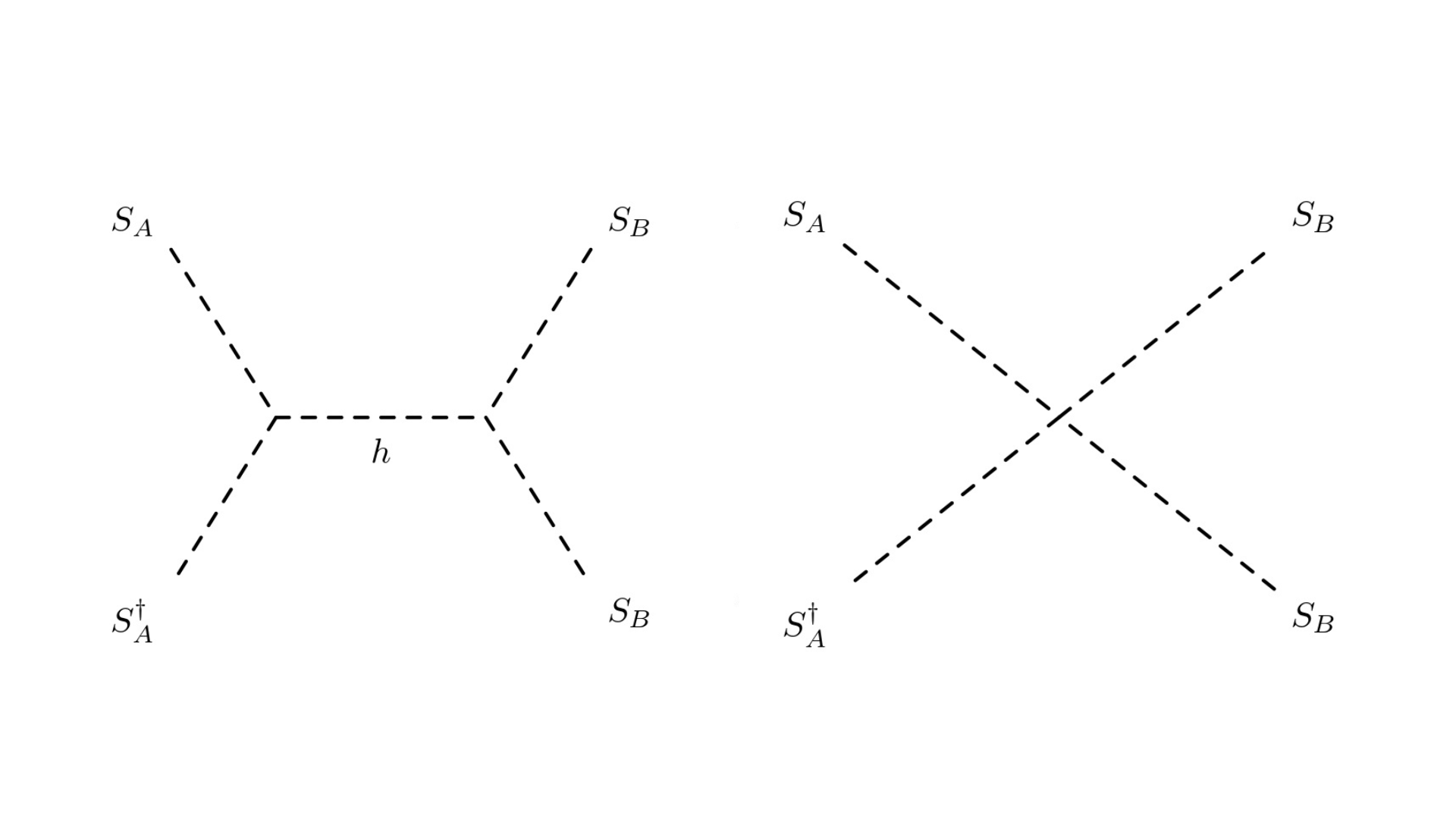}
        \caption{\justifying Dark-sector conversion,
        \(S_A S_A^\dagger \leftrightarrow S_B S_B\).}
        \label{fig:diagramZNconv}
    \end{subfigure}

    \vspace{0.25em}

    \begin{subfigure}{0.9\textwidth}
        \centering
        \includegraphics[width=\linewidth]{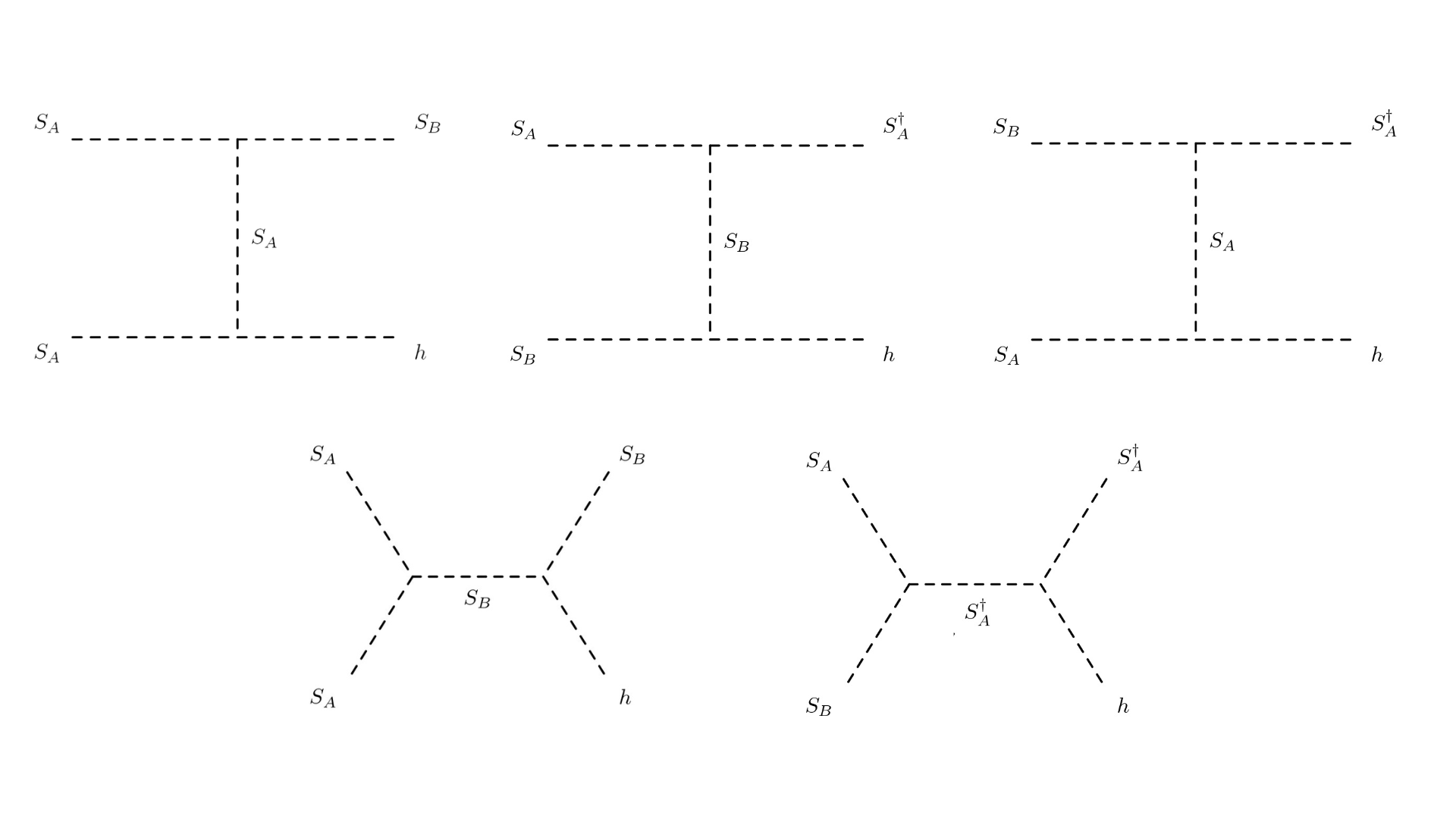}
        \caption{\justifying Representative \(\mu_{S1}\)-induced semi-annihilation,
        \(S_A S_A \to S_B h\), and its CP-conjugate process. The same interaction controls
        \(S_B\leftrightarrow S_A S_A\) when kinematically open.}
        \label{fig:diagramZ4}
    \end{subfigure}

    \caption{\justifying Key interaction topologies governing production, depletion, and redistribution of the two dark
    species in the coupled relic-density evolution.}
    \label{fig:diagramZ_all}
\end{figure}

The interactions introduced in Sec.~\ref{sec:model} generate the Higgs-portal annihilation, dark-sector conversion,
semi-annihilation, and decay topologies illustrated in Fig.~\ref{fig:diagramZ_all}. Their role in the relic-density
evolution depends on both the mass hierarchy and the thermal history of the two dark scalars. For
$M_{S_B}<2M_{S_A}$, both particles are stable and may contribute to the present dark matter abundance, whereas for
$M_{S_B}>2M_{S_A}$, the decay $S_B\to S_A S_A$ and its conjugate channel become kinematically allowed, leaving
$S_A$ as the final stable relic. We consider stable WIMP--FIMP and FIMP--FIMP regimes, together with
decay-mediated histories in which the unstable $S_B$ population contributes to the final $S_A$ abundance. The
corresponding production, depletion, and redistribution processes are incorporated into the coupled Boltzmann
system discussed below, while the parameter ranges associated with the different cosmological assignments are
specified in Sec.~\ref{sec:results}.

\subsection{Boltzmann Equations}
\label{subsec:relicdens}

To determine the relic abundance of the dark sector, we track the comoving number densities
\(Y_{S_i}=n_{S_i}/s\), where \(s\) is the entropy density, as a function of
\(x=M_{S_A}/T\). We assume vanishing particle--antiparticle asymmetries in the
dark sector. Thus, for the complex field \(S_A\), one has
\(Y_{S_A}=Y_{S_A^\dagger}\), and we denote both quantities by \(Y_{S_A}\), with
the same convention for the corresponding equilibrium yields. Accordingly,
when converting the asymptotic yield into the relic density of the complex
scalar sector, both the \(S_A\) and \(S_A^\dagger\) populations are included.
In the reaction labels, however, we keep the particle content explicit.

The relic-density calculation is performed with the multicomponent routines
of \texttt{micrOMEGAs}~\cite{micromegas}, with the relevant matrix
elements generated by \texttt{CalcHEP}~\cite{calchep}. The computation
includes the full set of processes generated for the model and retained in
the numerical analysis. The equations below are therefore not intended as an exhaustive transcription of the
numerical reaction network. Instead, they display the structure of the coupled
evolution and the leading number-changing processes that control the relic
abundance in the cosmological regimes considered here. We write

\begin{align}
\frac{dY_{S_A}}{dx} =\;&
-\sqrt{\frac{\pi}{45}}\frac{M_{\rm Pl}\,g_*^{1/2}M_{S_A}}{x^2}
\Bigg[
\langle \sigma v \rangle_{S_AS_A^\dagger \to {\rm SM}}
\left(Y_{S_A}^2 - Y_{S_A}^{\rm eq\,2} \right)
\nonumber\\
&+
\langle \sigma v \rangle_{S_AS_A^\dagger \to S_BS_B}
\left(
Y_{S_A}^2
-
Y_{S_B}^2
\frac{Y_{S_A}^{\rm eq\,2}}{Y_{S_B}^{\rm eq\,2}}
\right)
+
2\,\langle \sigma v \rangle_{S_AS_A \to S_B h}
\left(
Y_{S_A}^2
-
Y_{S_A}^{\rm eq\,2}
\frac{Y_{S_B}}{Y_{S_B}^{\rm eq}}
\right)
\nonumber\\
&-
\frac{\langle \Gamma_{S_B\to S_AS_A} \rangle}{s}
\left(
Y_{S_B}
-
Y_{S_A}^{2}
\frac{Y_{S_B}^{\rm eq}}{Y_{S_A}^{\rm eq\,2}}
\right)
-
\frac{\langle \Gamma_{h\to S_A S_A^\dagger} \rangle}{s}
Y_h^{\rm eq}
\left(
1-\frac{Y_{S_A}^{2}}{Y_{S_A}^{\rm eq\,2}}
\right)
\Bigg],
\label{eq:BoltzzA_fix}
\end{align}

and

\begin{align}
\frac{dY_{S_B}}{dx} =\;&
-\sqrt{\frac{\pi}{45}}\frac{M_{\rm Pl}\,g_*^{1/2}M_{S_A}}{x^2}
\Bigg[
\langle \sigma v \rangle_{S_BS_B \to {\rm SM}}
\left(Y_{S_B}^2 - Y_{S_B}^{\rm eq\,2} \right)
\nonumber\\
&-
2\,\langle \sigma v \rangle_{S_AS_A^\dagger \to S_BS_B}
\left(
Y_{S_A}^2
-
Y_{S_B}^2
\frac{Y_{S_A}^{\rm eq\,2}}{Y_{S_B}^{\rm eq\,2}}
\right)
-
2\,\langle \sigma v \rangle_{S_AS_A \to S_B h}
\left(
Y_{S_A}^2
-
Y_{S_A}^{\rm eq\,2}
\frac{Y_{S_B}}{Y_{S_B}^{\rm eq}}
\right)
\nonumber\\
&+
\frac{\langle \Gamma_{S_B\to S_AS_A} \rangle}{s}
\left(
Y_{S_B}
-
Y_{S_A}^{2}
\frac{Y_{S_B}^{\rm eq}}{Y_{S_A}^{\rm eq\,2}}
\right)
+
\langle \sigma v \rangle^{\rm eff}_{S_BS_A \to S_A^\dagger h}\,
Y_{S_A}
\left(Y_{S_B}-Y_{S_B}^{\rm eq}\right)
\nonumber\\
&-
2\,\frac{\langle \Gamma_{h\to S_BS_B} \rangle}{s}
Y_h^{\rm eq}
\left(
1-\frac{Y_{S_B}^{2}}{Y_{S_B}^{\rm eq\,2}}
\right)
\Bigg].
\label{eq:BoltzzB_fix}
\end{align}

Here \(M_{\rm Pl}\) is the non-reduced Planck mass, \(H\) is the Hubble
rate, and \(g_*^{1/2}\) denotes the standard effective combination of
relativistic degrees of freedom. The common prefactor in
Eqs.~\eqref{eq:BoltzzA_fix} and~\eqref{eq:BoltzzB_fix} can then be written as
\begin{equation}
\sqrt{\frac{\pi}{45}}\frac{M_{\rm Pl}\,g_*^{1/2}M_{S_A}}{x^2}
=
\frac{s}{Hx}.
\end{equation}

Therefore the \(2\to2\) terms appear with thermally averaged cross sections,
while the decay and inverse-decay terms enter the square brackets as
\(\langle\Gamma\rangle/s\).

The thermally averaged dark-sector decay rate is
\begin{equation}
\label{eq:GammaDef_fix}
\langle \Gamma_{S_B\to S_A S_A} \rangle
\equiv
\left\langle
\Gamma_{S_B\to S_A S_A}
+
\Gamma_{S_B\to S_A^\dagger S_A^\dagger}
\right\rangle
=
\Gamma_{S_B}^{\rm tot}\,
\frac{K_1(x_{S_B})}{K_2(x_{S_B})},
\end{equation}
with
\begin{equation}
x_{S_B}=\frac{M_{S_B}}{T}
=
\frac{M_{S_B}}{M_{S_A}}x.
\end{equation}

Here \(K_n\) is the modified Bessel function of the second kind. In the
CP-symmetric limit,
\(\Gamma_{S_B}^{\rm tot}=2\Gamma_{S_B\to S_AS_A}\).

The Higgs decay rates are treated in the same notation,

\begin{equation}
\label{eq:HiggsThermalDecayRates}
\langle \Gamma_{h\to S_iS_i^{(\dagger)}} \rangle
=
\Gamma_{h\to S_iS_i^{(\dagger)}}\,
\frac{K_1(x_h)}{K_2(x_h)},
\qquad
x_h=\frac{M_h}{T}
=
\frac{M_h}{M_{S_A}}x.
\end{equation}

The corresponding terms in Eqs.~\eqref{eq:BoltzzA_fix}
and~\eqref{eq:BoltzzB_fix} are proportional to the equilibrium Higgs yield
\(Y_h^{\rm eq}=n_h^{\rm eq}/s\), since the Higgs belongs to the thermal bath.
They are included only when the decay channels are kinematically open. The
factor of \(2\) multiplying the \(h\to S_BS_B\) contribution accounts for the
two identical \(S_B\) particles produced per Higgs decay.

For the conversion contribution we use
\(\langle \sigma v\rangle_{S_AS_A^\dagger\to S_BS_B}\); the inverse process is
encoded by the equilibrium-subtracted structure. For semi-annihilation,
\(\langle\sigma v\rangle_{S_AS_A\to S_Bh}\) denotes a single channel. The factor
of \(2\) in Eq.~\eqref{eq:BoltzzA_fix} accounts for the two \(S_A\) quanta
depleted in \(S_AS_A\to S_Bh\), while the factor of \(2\) in
Eq.~\eqref{eq:BoltzzB_fix} accounts for the CP-conjugate pair of channels
\(S_AS_A\to S_Bh\) and \(S_A^\dagger S_A^\dagger\to S_Bh\).

The effective rate in Eq.~\eqref{eq:BoltzzB_fix} is defined as
\begin{equation}
\langle \sigma v \rangle^{\rm eff}_{S_BS_A\to S_A^\dagger h}
\equiv
\langle \sigma v \rangle_{S_BS_A\to S_A^\dagger h}
+
\langle \sigma v \rangle_{S_BS_A^\dagger\to S_A h}.
\end{equation}

In the symmetric limit \(Y_{S_A}=Y_{S_A^\dagger}\), this topology exchanges
\(S_A\leftrightarrow S_A^\dagger\) and therefore does not change the total
\(S_A\) abundance, while it changes the \(S_B\) abundance.

For a FIMP component, the abundance remains far below equilibrium during
production. In this limit, the equilibrium-subtracted source terms reproduce the
usual freeze-in behaviour, while the full coupled system also captures
conversion, semi-annihilation, and dark-sector injection effects.

\subsection{Direct Detection Constraints}
\label{subsec:direct}

Direct-detection experiments probe the WIMP-like regions of parameter space through
spin-independent (SI) scattering off nuclei mediated by Higgs exchange. At zero
momentum transfer, the SI DM--nucleon cross section can be approximated by
\begin{equation}
\label{eq:SIDirectDetec}
\sigma_{S_i}^{\rm SI}
\simeq
\frac{f_N^2}{4\pi}\,
\frac{\mu_{iN}^{2}m_N^2}{M_h^4M_{S_i}^2}
\left(
\frac{g_{hS_iS_i^{(\dagger)}}}{v}
\right)^2,
\end{equation}
where \(\mu_{iN}=M_{S_i}m_N/(M_{S_i}+m_N)\) is the DM--nucleon reduced mass and
we take \(f_N\simeq 0.30\) for the scalar nucleon form factor. The coupling
\(g_{hS_iS_i^{(\dagger)}}\) denotes the trilinear Higgs--DM coupling entering
the scattering amplitude. With the scalar potential in Eq.~\eqref{eq:pot}, the
relevant interaction terms in the broken phase are
\begin{equation}
\mathcal L \supset
- g_{hS_A S_A^\dagger}\,hS_A^\dagger S_A
-\frac{1}{2}g_{hS_BS_B}\,hS_B^2,
\end{equation}
with
\begin{equation}
g_{hS_A S_A^\dagger}=\lambda_{HA}\,v,
\qquad
g_{hS_BS_B}=\lambda_{HB}\,v.
\end{equation}

In a multi-component scenario the predicted event rate scales with the local
number density of each species. Assuming no spatial segregation, we approximate
the local fractions by the cosmological relic-density fractions,
\begin{equation}
\xi_{S_i}\equiv
\frac{\Omega_{S_i}h^2}{\Omega_{\rm tot}h^2},
\qquad
\Omega_{\rm tot}h^2
=
\Omega_{S_A}h^2+\Omega_{S_B}h^2.
\end{equation}

We apply the LZ spin-independent bound~\cite{LZ:2025PRL_4p2ty} using the commonly adopted
approximate multi-component rescaling,
\begin{equation}
\label{eq:SIDirectDetecExc}
\sum_{i=A,B}
\frac{\xi_{S_i}\sigma_{S_i}^{\rm SI}}
{\sigma_{\rm limit}(M_{S_i})}
<1,
\end{equation}
where \(\sigma_{\rm limit}(M_{S_i})\) is the single-component 90\% C.L. upper limit at mass \(M_{S_i}\).

In the FIMP-like regimes considered below, the Higgs-portal coupling of the feebly interacting component is extremely small, so its contribution to the direct-detection rate is negligible. Direct detection therefore mainly constrains the WIMP-like component.

The relic-fraction rescaling in Eq.~\eqref{eq:SIDirectDetecExc} does not generically remove Higgs-portal direct-detection constraints. For a thermal Higgs-portal component away from resonances, thresholds, or additional depletion
channels, the same portal coupling controls both the freeze-out annihilation rate and the SI scattering cross section. Parametrically,
\begin{equation}
\sigma_i^{\rm SI}\propto \lambda_{Hi}^2,
\qquad
\Omega_i h^2\propto \langle\sigma v\rangle_i^{-1}
\sim \lambda_{Hi}^{-2},
\end{equation}
so that \(\xi_i\sigma_i^{\rm SI}\) is not arbitrarily reduced by increasing the
portal coupling. The WIMP-like component in the mixed scenarios therefore
remains subject to the usual Higgs-portal correlation between freeze-out and
direct detection. Viable thermal points are expected mainly near the Higgs
resonance or in regions where additional dark-sector processes modify the
standard freeze-out evolution.

For completeness, we do not impose indirect-detection constraints in this work. In the FIMP-like regimes,
present-day annihilation signals into SM states are strongly suppressed by the feeble portal interactions. For a WIMP-like subcomponent,
annihilation rates into SM final states are suppressed by the square of its local density fraction, \(\xi_i^2\).
In addition, semi-annihilation channels such as \(S_A S_A\to S_B h\) may contribute when kinematically open. Since
the resulting spectra depend on the mass hierarchy, branching fractions, and relic-density fractions, they are not
directly described by the standard single-component limits. A dedicated recast would therefore be required, which
we leave for future work.

\subsection{Higgs Invisible Decay}
\label{subsec:higgs}

The Higgs portal provides an additional collider constraint whenever one of
the dark scalars is light enough to be produced in Higgs decays. For
\(M_{S_i}<M_h/2\), the corresponding invisible channels are
\[
h\to S_A S_A^\dagger,
\qquad
h\to S_BS_B.
\]

Their partial widths follow directly from the trilinear Higgs--DM couplings
defined in Sec.~\ref{subsec:direct},
\[
\lambda_{HA}=\frac{g_{hS_A S_A^\dagger}}{v},
\qquad
\lambda_{HB}=\frac{g_{hS_BS_B}}{v}.
\]
Defining
\(\Gamma_A\equiv\Gamma_{h\to S_A S_A^\dagger}\) and
\(\Gamma_B\equiv\Gamma_{h\to S_BS_B}\), the two partial widths can be written
compactly as
\begin{equation}
\Gamma_i
=
\kappa_i\,
\frac{\lambda_{Hi}^2v^2}{16\pi M_h}
\left(1-\frac{4M_{S_i}^2}{M_h^2}\right)^{1/2},
\qquad
\kappa_A=1,
\qquad
\kappa_B=\frac{1}{2},
\end{equation}

where the factor \(\kappa_B=1/2\) accounts for the two identical real
scalars in the \(S_BS_B\) final state, while \(S_A\) and \(S_A^\dagger\)
are distinct particles.

The total invisible width \(\Gamma_{\rm inv}\) is obtained by summing the
kinematically open channels and determines the invisible branching ratio,
\begin{equation}
\beta_{\rm inv}
\equiv
{\rm BR}(h\to {\rm inv})
=
\frac{\Gamma_{\rm inv}}
{\Gamma_h^{\rm SM}+\Gamma_{\rm inv}}.
\end{equation}

We impose the ATLAS Run~1 and Run~2 combination bound
\[
\beta_{\rm inv}\leq 0.107\;(0.077)
\]
at 95\% C.L. observed (expected)~\cite{ATLAS:2023InvComb}.

The phenomenological impact of this constraint depends directly on the
portal strength. It is most relevant for WIMP-like states with sizeable
Higgs couplings, while the feeble portals characteristic of the FIMP regime
strongly suppress the corresponding invisible contribution.

\subsection{Cosmological Constraints}
\label{subsec:cosmoconstraints}

It is important to clarify the cosmological status of the decay
\(S_B\to S_A S_A\), relevant in the hierarchy \(M_{S_B}>2M_{S_A}\).
Although the decay proceeds entirely within the dark sector, two indirect
cosmological effects can remain: a metastable \(S_B\) population contributes
to the total energy density and may modify the expansion rate near BBN, while
the \(S_A\) particles produced in late decays are born with a non-thermal
momentum and may free stream, suppressing small-scale structure.

In this work, for each viable point satisfying the theoretical,
relic-density, applicable Higgs invisible-decay, direct-detection, and
thermalisation requirements, we evaluate the lifetime \(\tau_{S_B}\), the
energy fraction \(r_{\rm BBN}\) carried by \(S_B\) at decay, the fraction
\(f_{\rm dec}\) of the final \(S_A\) abundance supplied during the
decay-injection epoch, and an equivalent thermal-WDM mass
\(m_{\rm WDM}^{\rm eq}\) for the injected \(S_A\) population. These
quantities are used to assess the possible impact of the decay-mediated
histories on BBN and small-scale structure.

\subsubsection{Big Bang Nucleosynthesis}

BBN bounds on long-lived particles are most severe when the decays inject
energetic photons or hadrons into the primordial plasma, photodissociating or
hadrodissociating the light nuclei and altering the predicted abundances of
\(\mathrm{D}\), \(^4\mathrm{He}\) and \(^7\mathrm{Li}\)
(see, e.g., Refs.~\cite{Kawasaki:2005,Jedamzik:2006,Kawasaki:2018,Cyburt:2016}).
In the present \(\mathbb{Z}_4\) model, however, the decay \(S_B\to S_A S_A\)
and its CP-conjugate contain no SM particles in the final state. The usual
electromagnetic and hadronic BBN bounds on visible decay products of the
metastable parent therefore do not apply directly, and we do not impose them.

The total dark-sector decay width, summing the two CP-conjugate channels, and the lifetime of $S_{B}$ are
\begin{equation}
\label{eq:GammaSB_tot}
\Gamma_{S_B}^{\rm tot}
\simeq
\frac{|\mu_{S1}|^2}{16\pi M_{S_B}}
\left(1-\frac{4M_{S_A}^2}{M_{S_B}^2}\right)^{1/2},
\qquad
\tau_{S_B}=\big(\Gamma_{S_B}^{\rm tot}\big)^{-1},
\end{equation}
where \(\mu_{S1}\) is the trilinear coupling of Eq.~\eqref{eq:pot}, \(M_{S_B}\)
is the parent mass, and the square root is the two-body phase-space factor that
vanishes at the kinematic threshold \(M_{S_B}=2M_{S_A}\). Over the trilinear
range of the scan, \(\tau_{S_B}\) spans many orders of magnitude: for the feeble
end (\(|\mu_{S1}|\sim10^{-12}\,\mathrm{GeV}\)) the lifetime can substantially
exceed \(1\,\mathrm{s}\), so we do not rely on an ``early decay'' assumption.
We therefore focus on the possible effect of the metastable parent population
on the expansion rate.

To estimate this effect, we use the energy density of the metastable \(S_B\)
population relative to radiation. We adopt the sudden-decay approximation and
define an effective decay temperature \(T_d\) through
\(\Gamma_{S_B}^{\rm tot}=H(T_d)\), with
\(H=\sqrt{\frac{4\pi^{3}\,g_*}{45}}\frac{T^2}{M_{\rm Pl}}\). For the retained
decay-mediated points, \(T_d\ll M_{S_B}\), so the parent is non-relativistic
at the decay epoch. Writing the \(S_B\) number density as
\(n_{S_B}=Y_{S_B}^{\rm pre}\,s\), with the entropy density
\(s=\tfrac{2\pi^2}{45}h_{\rm eff}T^3\) and the radiation energy density
\(\rho_{\rm rad}=\tfrac{\pi^2}{30}g_{\rm eff}T^4\), the corresponding energy
fraction reads
\begin{equation}
\label{eq:rBBN}
r_{\rm BBN}
\equiv
\left.\frac{\rho_{S_B}}{\rho_{\rm rad}}\right|_{T_d}
\simeq
\frac{4}{3}\,\frac{h_{\rm eff}}{g_{\rm eff}}\,
\frac{M_{S_B}\,Y_{S_B}^{\rm pre}}{T_d}\, .
\end{equation}

Here \(Y_{S_B}^{\rm pre}=n_{S_B}/s\) is the frozen comoving \(S_B\) yield just
before decay, \(M_{S_B}Y_{S_B}^{\rm pre}\) is the comoving energy per entropy
carried by \(S_B\), \(T_d\) is the effective decay temperature, and
\(g_{\rm eff}\) and \(h_{\rm eff}\) are the effective relativistic degrees
of freedom for the energy and entropy densities
(\(h_{\rm eff}/g_{\rm eff}\simeq1\) at the temperatures of interest).
Since \(S_B\) is non-relativistic before decay,
\(\rho_{S_B}/\rho_{\rm rad}\propto T^{-1}\), so this ratio grows as the
Universe cools and reaches its largest pre-decay value at \(T_d\).
Consequently, \(r_{\rm BBN}\ll1\) provides a conservative criterion ensuring
that the metastable \(S_B\) population never makes an appreciable contribution
to the expansion rate prior to its decay. For all viable points retained in
our scan we find \(r_{\rm BBN}\ll1\), so no expansion-driven BBN constraint
arises from the parent population. A more complete treatment could translate
the extra dark-sector energy density during the BBN epoch into a corresponding
bound on the expansion rate, or equivalently an effective
\(\Delta N_{\rm eff}\)-type constraint~\cite{PospelovPradler:2010,Iocco:2009}. Given the smallness
of \(r_{\rm BBN}\), such a refinement would not affect our conclusions.

\subsubsection{Lyman-\texorpdfstring{$\alpha$}{alpha} Forest}

Dark matter produced in late decays carries a non-thermal momentum distribution
and may behave as a warm subcomponent, suppressing the matter power spectrum on
small scales. The Lyman-\(\alpha\) forest is sensitive to such suppression, and
the corresponding limits depend on the warm-component fraction, the decay time,
and the daughter momentum distribution~\cite{Viel:2013,Irsic:2017,Villasenor:2023,Decant:2022}. A first diagnostic is the fraction of the final \(S_A\) abundance generated
during the late decay-injection epoch. We define
\begin{equation}
\label{eq:fdec}
f_{\rm dec}
\equiv
\frac{Y_{S_A}^{\rm final}-Y_{S_A}^{\rm pre}}
{Y_{S_A}^{\rm final}},
\end{equation}
where \(Y_{S_A}^{\rm pre}\) is the pre-injection plateau value identified
numerically from the coupled evolution before the onset of the decay-induced
increase, while \(Y_{S_A}^{\rm final}\) is the asymptotic post-injection
yield. In the decay-mediated regimes considered here, the rise
between these two plateaus is dominated by \(S_B\to S_A S_A\) and its
conjugate channel, so \(f_{\rm dec}\) provides a direct estimate of the
fraction of the final abundance supplied by late dark-sector decays.

A sizeable \(f_{\rm dec}\) does not by itself imply an excluded warm component,
since the impact also depends on how relativistic the produced particles are.
Within the same sudden-decay approximation, we characterise the injected
population by the two-body momentum evaluated in the \(S_B\) rest frame and
redshifted from the effective decay temperature \(T_d\). Each \(S_A\) is
therefore assigned the characteristic production momentum
\begin{equation}
\label{eq:pstar}
p_\star=\frac{M_{S_B}}{2}\left(1-\frac{4M_{S_A}^2}{M_{S_B}^2}\right)^{1/2},
\end{equation}
in the \(S_B\) rest frame, which redshifts to a present-day velocity
\begin{equation}
\label{eq:v0}
v_0=
c\,\frac{p_\star}{M_{S_A}}\,\frac{a_d}{a_0}
=
c\,\frac{p_\star}{M_{S_A}}\,\frac{T_0}{T_d}
\left(\frac{g_{*s}(T_0)}{g_{*s}(T_d)}\right)^{1/3},
\end{equation}
where \(c\) is the speed of light, \(T_0\) is the present photon temperature,
\(T_d\) the effective decay temperature defined above, and \(a_d/a_0\) the
scale-factor ratio from decay to today. The \(g_{*s}\) factors account for
comoving entropy conservation. With the explicit factor of \(c\), \(v_0\)
is expressed in physical velocity units and can be directly compared with
the thermal-WDM relation below.

The factor \(p_\star/M_{S_A}\) characterises the daughter momentum at production,
while \(a_d/a_0\) accounts for its subsequent redshift. Accordingly, \(v_0\)
should be understood as a characteristic present-day velocity within the
sudden-decay approximation, rather than as a reconstruction of the complete
non-thermal phase-space distribution. For the warmness estimate, we set the entropy-redshift factor
\(\left[g_{*s}(T_0)/g_{*s}(T_d)\right]^{1/3}\) to unity. This approximation
overestimates the present-day daughter velocity and therefore provides a
conservative warmness diagnostic.

To obtain a diagnostic comparison with published thermal-WDM bounds, we
define an equivalent thermal-relic mass \(m_{\rm WDM}^{\rm eq}\) by matching
the characteristic present-day velocity \(v_0\) to the present-day rms
velocity of a thermal relic. For a thermal
relic~\cite{BodeOstrikerTurok:2001,BarkanaHaimanOstriker:2001},
\begin{equation}
\label{eq:vrms}
v_{\rm rms}^{\rm WDM}
=0.0437(1+z)\left(\frac{\Omega_X h^2}{0.15}\right)^{1/3}
\left(\frac{g_X}{1.5}\right)^{-1/3}
\left(\frac{m_X}{\mathrm{keV}}\right)^{-4/3}\ \mathrm{km/s},
\end{equation}
where \(m_X\) is the thermal-WDM mass, \(g_X\) its internal degrees of
freedom, and \(\Omega_X h^2\) its relic density. Equating
\(v_{\rm rms}^{\rm WDM}\) at \(z=0\) to the characteristic present-day
velocity \(v_0\), and adopting
\((\Omega_X h^2/0.15)^{1/3}\simeq1\) and
\((g_X/1.5)^{-1/3}\simeq1\), gives
\begin{equation}
\label{eq:mwdmeq}
m_{\rm WDM}^{\rm eq}
\simeq
\left(\frac{0.0437\,\mathrm{km/s}}{v_0}\right)^{3/4}\ \mathrm{keV},
\end{equation}
which defines the thermal-relic mass whose present-day rms velocity matches
the characteristic velocity \(v_0\) of the decay-produced \(S_A\)
population. Within this velocity-matching diagnostic, a smaller
\(m_{\rm WDM}^{\rm eq}\) corresponds to a warmer injected component.

As reference limits we consider the Lyman-\(\alpha\) bounds obtained for
thermal-relic WDM. An important systematic in these analyses is the assumed
thermal history of the intergalactic medium. The hydrodynamical study of
Ref.~\cite{Villasenor:2023}, which marginalises over a wide range of
physically motivated thermal histories using the small-scale flux power
spectrum measured at high redshift
(\(4.0\lesssim z\lesssim5.2\)), obtains
\(m_{\rm WDM}\gtrsim3.1~\mathrm{keV}\) (95\% C.L.); we adopt
\(3.1~\mathrm{keV}\) as our fiducial thermal-WDM reference value. This is
consistent with the earlier HIRES/MIKE analysis of
Ref.~\cite{Viel:2013}
(\(m_{\rm WDM}\gtrsim3.3~\mathrm{keV}\), \(2\sigma\)) and lies below the
combined XQ-100\,+\,HIRES/MIKE result of Ref.~\cite{Irsic:2017},
\(m_{\rm WDM}\gtrsim5.3~\mathrm{keV}\) at \(2\sigma\), which relaxes to
\(\simeq3.5~\mathrm{keV}\) for a less restrictive thermal history. More
conservative treatments allow masses as low as
\(\simeq1.9~\mathrm{keV}\)~\cite{Garzilli:2021}. We therefore use
\(3.1<m_{\rm WDM}<5.3~\mathrm{keV}\) as a reference interval illustrating
the spread among representative thermal-WDM limits, while
\(3.1~\mathrm{keV}\) sets the threshold used in our warmness flag.

Two caveats must be stressed. First, these bounds are derived for a
\emph{thermal} Fermi--Dirac relic, whereas the population injected by
\(S_B\) decay is non-thermal. The velocity matching of
Eq.~\eqref{eq:mwdmeq} therefore provides a simple warmness diagnostic but
does not reproduce the detailed shape of the power-spectrum cutoff. Second,
only the fraction \(f_{\rm dec}\) of the final abundance is associated with
the decay-injection epoch. For \(f_{\rm dec}<1\), the corresponding
structure-formation constraint is generally expected to be weaker than the
pure-WDM reference limit, as also found in dedicated analyses of mixed
cold--warm dark matter scenarios~\cite{GarciaGallego:2025CWDM}. Those
analyses, however, assume a thermal warm component and therefore cannot be
transferred directly to the non-thermal population produced by \(S_B\)
decays. A rigorous constraint in our scenario would require the full
non-thermal phase-space distribution of the stable particle and the resulting
matter transfer function.

For these reasons, we use \(m_{\rm WDM}^{\rm eq}\) together with
\(f_{\rm dec}\) as a diagnostic rather than as a hard exclusion. Operationally, we flag configurations with \(f_{\rm dec}>0.1\) and
\(m_{\rm WDM}^{\rm eq}<3.1~\mathrm{keV}\) as potentially warm. Equivalent masses in the
interval \(3.1<m_{\rm WDM}^{\rm eq}<5.3~\mathrm{keV}\) lie within the range
spanned by representative thermal-WDM limits and should therefore be regarded
as sensitive to the adopted thermal benchmark rather than as a
model-independent exclusion. Configurations with
\(m_{\rm WDM}^{\rm eq}>5.3~\mathrm{keV}\), or with
\(f_{\rm dec}<0.1\), are not flagged as potentially warm by this
criterion. In all cases, a quantitative structure-formation bound would
require a dedicated calculation of the daughter momentum distribution and
its effect on the matter power spectrum. We therefore retain the flagged
points in the present analysis and leave such a treatment for future work.

\section{Numerical Analysis and Results}
\label{sec:results}

In this section we present the numerical analysis of the viable parameter space
of the \(\mathbb{Z}_4\) model, focusing on the interplay between thermal
freeze-out and non-thermal freeze-in production. Standard-Model quantities are
set to their PDG values~\cite{PDGDM}, while
\(\lambda_A\), \(\lambda_B\), and \(\lambda_{S4}\) are fixed to representative
benchmark values. These self-couplings do not directly control the leading production and
depletion channels relevant to the cosmological regimes considered here,
while theoretical consistency is imposed point by point through the
boundedness-from-below, perturbativity, and perturbative-unitarity conditions
described in Sec.~\ref{sec:model} and Appendix~\ref{app:unitarity}.

The scanned parameters and their ranges are summarised in
Table~\ref{tab:scan_params}.  We take
\(40~{\rm GeV}\leq M_{S_A}\leq2~{\rm TeV}\). The range of \(M_{S_B}\)
depends on the mass hierarchy: \(M_{S_B}<2M_{S_A}\) in the stable
two-component regime, while \(2M_{S_A}<M_{S_B}\leq4~{\rm TeV}\) in the
decay-mediated regime.

Following the WIMP-like/FIMP-like terminology introduced in Sec.~\ref{sec:cosmo},
we use different Higgs-portal ranges to target the corresponding thermal
regimes. For a WIMP-like component \(S_i\), with \(i=A,B\), we scan
\(|\lambda_{Hi}|\in[10^{-4},1]\), whereas for a FIMP-like component we take
\(|\lambda_{Hi}|\in[10^{-12},10^{-8}]\), as in
Refs.~\cite{Pukhov,YagunaZapata2024}. Here \(\lambda_{Hi}\) denotes
\(\lambda_{HA}\) for \(S_A\) and \(\lambda_{HB}\) for \(S_B\). The magnitudes
of the scanned dimensionless couplings are sampled logarithmically, with their
signs chosen independently. The mass parameters are explored through a
combination of broad logarithmic sampling and targeted local scans in regions
requiring finer resolution, in particular around the Higgs resonance.

For Higgs-mediated observables, namely direct detection and invisible Higgs
decays, we use the broken-phase couplings
\(g_{hS_iS_i^{(\dagger)}}\) defined in Sec.~\ref{subsec:direct}. The
intra-dark interactions are scanned over
\(|\lambda_{AB}|\in[10^{-12},10^{-8}]\) and
\(|\mu_{S1}|\in[10^{-12},10^{-1}]~{\rm GeV}\), allowing conversion,
semi-annihilation, and decay-mediated injection while targeting the
non-thermal regimes of interest. The WIMP/FIMP assignment is then verified
a posteriori through the thermalisation criterion described in
Appendix~\ref{subsec:thermalisation_check}, by comparing the relevant
interaction rates with the Hubble expansion rate over the temperature grid
used to reconstruct the thermal history.

For each scenario we generate samples of order \(\mathcal O(10^7)\) trial points.
The numerical calculations are performed with \texttt{micrOMEGAs}~6.0, using its multicomponent routines to solve the
coupled Boltzmann system. The relic-density requirement is imposed on the
total asymptotic abundance,
\[
\Omega_{\rm tot}h^2=\Omega_{S_A}h^2+\Omega_{S_B}h^2,
\]
where \(\Omega_{S_A}h^2\) includes the combined \(S_A\) and
\(S_A^\dagger\) populations. In the stable regime, both dark-sector
components contribute, while \(\Omega_{S_B}h^2\to0\) asymptotically in the
decay-mediated regime. The total abundance must agree with the Planck 2018
determination,
\(\Omega_{\rm Planck}h^2=0.1200\pm0.0012\) at \(68\%\) C.L.,
within a \(2\sigma\) experimental interval enlarged by a conservative
\(10\%\) relative theoretical allowance on the relic-density calculation,
added in quadrature,
\begin{equation}
\label{eq:omega_criterion}
\left|\Omega_{\rm tot} h^2-\Omega_{\rm Planck}h^2\right|
\leq
\sqrt{
\left(2\sigma_{\rm Planck}\right)^2
+
\left(0.1\,\Omega_{\rm Planck}h^2\right)^2
}\, .
\end{equation}

Within the set of constraints considered in this work, we refer to a point
as viable when, in addition to satisfying this relic-density criterion, it
fulfills the theoretical consistency conditions discussed in
Sec.~\ref{sec:model} and Appendix~\ref{app:unitarity}, the applicable Higgs
invisible-decay constraint, the LZ spin-independent direct-detection
bound~\cite{LZ:2025PRL_4p2ty}, and the thermalisation requirements described
in Appendix~\ref{subsec:thermalisation_check}.

\begin{table}[t]
\centering
\renewcommand{\arraystretch}{1.3}
\setlength{\tabcolsep}{8pt}
\caption{\justifying Parameter ranges adopted in the numerical scan. For the
scanned dimensionless couplings, the ranges refer to absolute values; their signs
are sampled independently. The Higgs-portal range depends on the assumed thermal
nature of the corresponding component.}
\label{tab:scan_params}
\begin{tabular}{@{}llp{6.5cm}@{}}
\toprule
\textbf{Parameter} & \textbf{Scanning Range} & \textbf{Physical Role} \\
\midrule
\multicolumn{3}{l}{\itshape Mass parameters} \\
\(M_{S_A}\) & \([40,\,2000]\) GeV & Stable DM mass \\
\(M_{S_B}\) &
\(\begin{array}{l}
[40,\,2M_{S_A})~{\rm GeV}\\[-0.15em]
\hspace{1em}\text{(stable)}\\
(2M_{S_A},\,4000]~{\rm GeV}\\[-0.15em]
\hspace{1em}\text{(decay-mediated)}
\end{array}\)
& DM or parent mass \\
\midrule
\multicolumn{3}{l}{\itshape Higgs portal couplings} \\
\(|\lambda_{HA}|\) or \(|\lambda_{HB}|\) for WIMP-like \(S_i\)
& \([10^{-4},\,1]\) & Thermal contact with the SM bath \\
\(|\lambda_{HA}|\) or \(|\lambda_{HB}|\) for FIMP-like \(S_i\)
& \([10^{-12},\,10^{-8}]\) & Feeble contact with the SM bath \\
\midrule
\multicolumn{3}{l}{\itshape Dark-sector couplings} \\
\(|\lambda_{AB}|\) & \([10^{-12},\,10^{-8}]\) & Dark-sector conversion \\
\(|\mu_{S1}|\) & \([10^{-12},\,10^{-1}]\) GeV & Semi-annihilation and decay \\
\midrule
\multicolumn{3}{l}{\itshape Fixed parameters} \\
\(\lambda_A,\lambda_B,\lambda_{S4}\)
& \((0.03,\;0.02,\;0.01)\) & Scalar self-interactions and vacuum structure \\
\bottomrule
\end{tabular}
\end{table}

The mass hierarchy and the thermal history of the two scalars define six
benchmark scenarios, summarised in Table~\ref{tab:scenarios}. Scenarios~1--3
correspond to the stable regime, \(M_{S_B}<2M_{S_A}\), in which both scalars
survive and may contribute to the present dark matter abundance. Scenario~1 is
the fully non-thermal FIMP--FIMP case, while Scenarios~2 and~3 contain one
WIMP-like and one FIMP-like component with their roles interchanged.

Scenarios~4--6 correspond to the decay-mediated regime,
\(M_{S_B}>2M_{S_A}\), where \(S_B\to S_A S_A\) is kinematically open and
only \(S_A\) survives asymptotically. They realise, respectively, a SuperWIMP
history with a thermal parent, injection into a WIMP-like daughter, and
sequential freeze-in with both states feebly coupled. The detailed
phenomenology of these six cases is analysed below.

\begin{table}[!htbp]
\centering
\small
\renewcommand{\arraystretch}{1.08}
\setlength{\tabcolsep}{4pt}
\caption{\justifying Classification of the six cosmological scenarios considered
in the scan. The ``Nature'' column gives the WIMP/FIMP assignment of
\((S_A,S_B)\); ``Stability'' indicates whether \(S_B\to S_A S_A\) is open.}
\label{tab:scenarios}

\newcolumntype{Y}{>{\RaggedRight\arraybackslash}X}

\begin{tabularx}{\linewidth}{@{}c l c l Y@{}}
\toprule
\textbf{Sc.} & \textbf{Nature \((S_A,S_B)\)} & \textbf{Stability} & \textbf{Mechanism} & \textbf{Physical feature} \\
\midrule
\multicolumn{5}{@{}l}{\itshape Stable regime: \(M_{S_B}<2M_{S_A}\)} \\

\textbf{1} & FIMP + FIMP & Stable & Pure freeze-in &
Two stable FIMPs whose combined freeze-in production determines the relic
abundance. \\

\textbf{2} & FIMP + WIMP & Stable & Mixed &
Thermal \(S_B\) plus feebly coupled \(S_A\); \(S_B\) is tested by the
direct-detection rate rescaled by its relic fraction, while \(S_A\) can
supply the remaining abundance. \\

\textbf{3} & WIMP + FIMP & Stable & Mixed &
Thermal \(S_A\) plus feebly coupled \(S_B\); \(S_A\) is tested by the
relic-fraction-rescaled direct-detection rate, while \(S_B\) supplies a
non-thermal contribution. \\

\midrule
\multicolumn{5}{@{}l}{\itshape Decay-mediated regime: \(M_{S_B}>2M_{S_A}\)} \\

\textbf{4} & FIMP + WIMP & \(S_B\to S_A S_A\) & SuperWIMP &
WIMP-like \(S_B\) freezes out and later decays into feebly coupled \(S_A\). \\

\textbf{5} & WIMP + FIMP & \(S_B\to S_A S_A\) & Injection &
FIMP-like \(S_B\) injects \(S_A\), adding a non-thermal contribution to the
WIMP-like daughter abundance. \\

\textbf{6} & FIMP + FIMP & \(S_B\to S_A S_A\) & Sequential freeze-in &
Both sectors are feeble; \(S_A\) receives contributions from direct freeze-in
production and subsequent \(S_B\) decay. \\
\bottomrule
\end{tabularx}
\end{table}

\subsection{Stable Scenarios}
\label{subsec:stablescenarios}

We begin with the stable configurations, defined by the hierarchy
\(M_{S_B}<2M_{S_A}\). In this regime the two-body decay
\(S_B\to S_A S_A\) is kinematically forbidden, so both species can contribute
to the present dark matter abundance. We analyse first the fully feeble
FIMP--FIMP case and then the two mixed configurations, in which either
\(S_B\) or \(S_A\) is the WIMP-like component.

\subsubsection{Pure Freeze-in: the FIMP--FIMP Regime}
\label{subsec:purefimpscenario}

We first consider the scenario in which both dark scalars behave as feebly
interacting massive particles, as in Ref.~\cite{Pandey:2018FIMP}. In this
regime the Higgs-portal magnitudes satisfy
\(|\lambda_{HA}|,|\lambda_{HB}|\lesssim10^{-8}\) within our scan ranges, and
the accepted points remain out of thermal equilibrium with the SM bath
according to the thermalisation check described in
Appendix~\ref{subsec:thermalisation_check}. The relic abundances of
\(S_A\) and \(S_B\) are therefore generated through freeze-in.

For the viable points in this regime, the strongly suppressed Higgs portals
render the direct-detection rates negligible. After imposing the theoretical
and thermalisation requirements, the dominant phenomenological selection is
therefore the total relic abundance and its partition between \(S_A\) and
\(S_B\).

\begin{figure}[htbp]
    \centering
    \begin{subfigure}{0.48\textwidth}
        \centering
        \includegraphics[width=\textwidth]{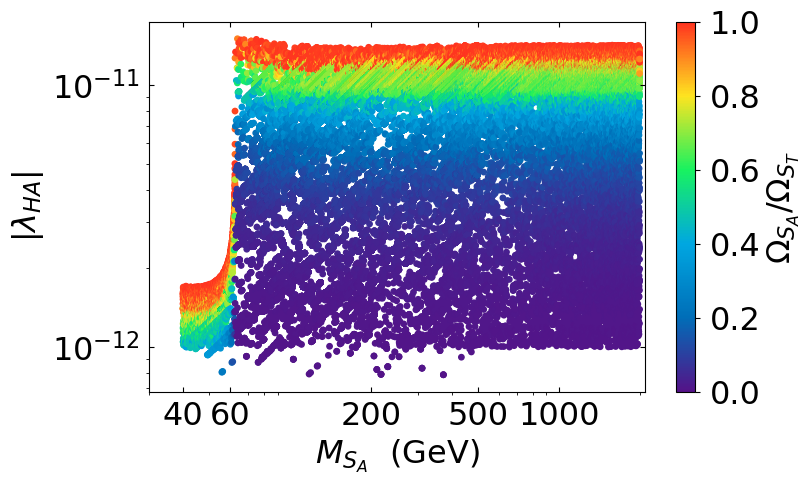}
    \end{subfigure}
    \hfill
    \begin{subfigure}{0.48\textwidth}
        \centering
        \includegraphics[width=\textwidth]{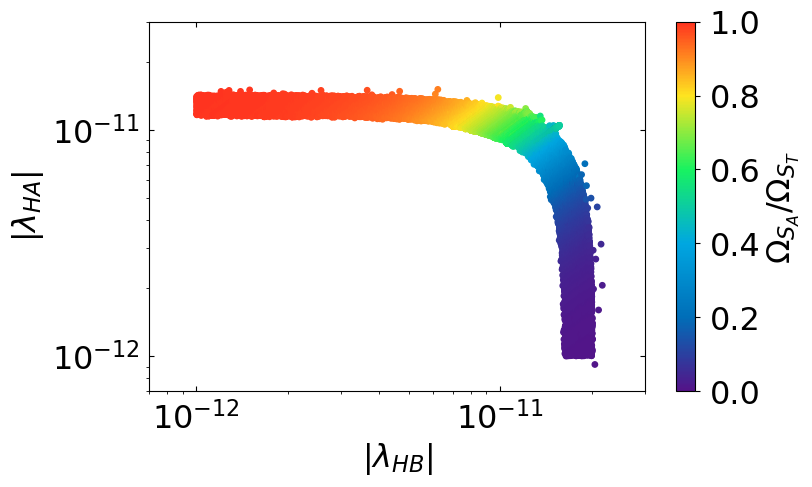}
    \end{subfigure}
   \caption{\justifying Parameter space for the FIMP--FIMP regime.
\textbf{Left:} Freeze-in scaling for \(S_A\) in the limit where it dominates
the total relic abundance, and its deformation as the \(S_B\) contribution
becomes non-negligible.
\textbf{Right:} Correlation between the portal-coupling magnitudes
\(|\lambda_{HA}|\) and \(|\lambda_{HB}|\), colour-coded by the fractional
abundance. The curved band identifies the region in which the two freeze-in
contributions combine to reproduce the observed total relic density.}
    \label{fig:Z4FFS}
\end{figure}

The viable parameter space is shown in Fig.~\ref{fig:Z4FFS}. The left panel
shows that the single-component freeze-in behaviour for \(S_A\) is recovered
when it dominates the total abundance,
\(\Omega_{S_A}/\Omega_{\rm tot}\to1\), in agreement with the
trends found in single-FIMP studies (see, e.g., Fig.~6 of
Ref.~\cite{Yaguna:2011FIMP}). In this regime, the two abundances are
predominantly generated through their respective freeze-in channels, while
their final partition is fixed by the requirement that the combined abundance
reproduce the observed relic density. Accordingly, solutions with comparable
contributions, \(\Omega_{S_A}\simeq\Omega_{S_B}\), tend to exhibit similar
portal-coupling magnitudes, as shown in the right panel. The leading production
rates scale quadratically with the corresponding portal couplings, while the
masses and kinematic regimes determine the detailed relation between the two
abundances. This balance produces the curved band in the
\((|\lambda_{HA}|,|\lambda_{HB}|)\) plane, similar to the feature identified
in Ref.~\cite{Pandey:2018FIMP} for two-component freeze-in.

This behaviour differs from the two-WIMP realisations explored in
\(\mathbb{Z}_{2n}\)-stabilised scenarios. In the scans of
Refs.~\cite{CarvalhoCorrea:2025Z2n,Yaguna2021}, where both components
thermalise with the SM bath and undergo freeze-out, viable regions are often
characterised by one component providing the dominant relic fraction while
the other remains subdominant
(see, e.g., Fig.~4 of Ref.~\cite{CarvalhoCorrea:2025Z2n} and Figs.~4 and~5 of
Ref.~\cite{Yaguna2021}). In the present FIMP--FIMP scan, by contrast, viable
solutions also occur in which \(S_A\) and \(S_B\) provide comparable fractions
of the observed dark matter abundance. This illustrates that a more balanced relic-density partition is possible when both components are produced through freeze-in.

Consequently, parameter regions that would be underabundant if only one component were included can reproduce the observed relic density once the second FIMP contribution is taken into account. The resulting correlation between the two portal couplings reflects the requirement that their combined freeze-in contributions match the observed total abundance, with the allowed combinations largely selected by the relic-density constraint.

\subsubsection{Mixed stable scenario with a \texorpdfstring{\(S_B\)}{SB} WIMP component}
\label{subsubsec:fimpwimp}

In this mixed stable scenario, the real scalar \(S_B\) is the thermal component,
coupled to the SM through \(\lambda_{HB}\), while the complex scalar \(S_A\)
remains feebly coupled and is produced through freeze-in. The final relic-density
composition is determined by the coupled thermal/non-thermal evolution, so the
thermal component \(S_B\) is not required to saturate the observed dark matter
density by itself.

\begin{figure}[htbp]
    \centering
    \begin{subfigure}{1.0\textwidth}
        \centering
        \begin{minipage}{0.48\textwidth}
            \centering
            \includegraphics[width=\textwidth]{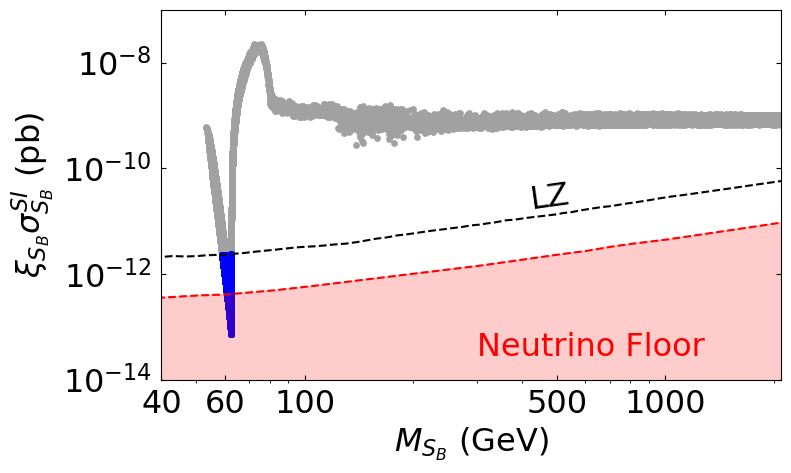}
        \end{minipage}
        \hfill
        \begin{minipage}{0.48\textwidth}
            \centering
            \includegraphics[width=\textwidth]{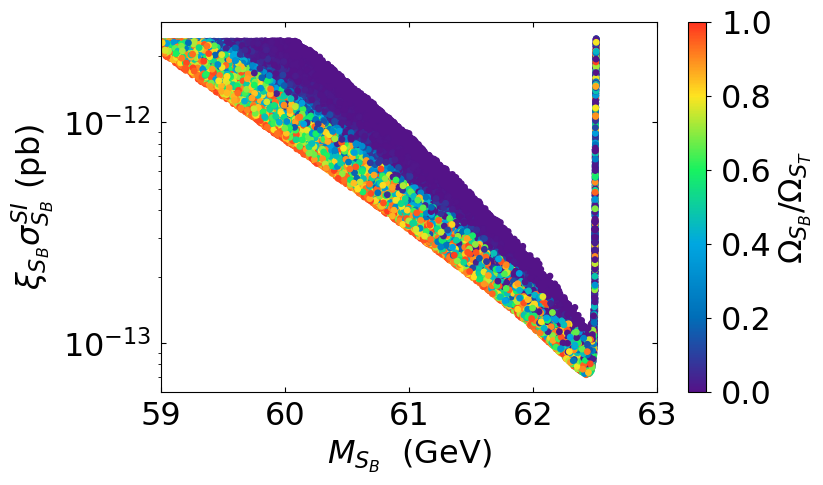}
        \end{minipage}
    \end{subfigure}
  \caption{\justifying Direct-detection prospects for the mixed stable scenario
with \(S_B\) as the WIMP-like component and \(S_A\) as the FIMP-like
component. All points shown satisfy the Planck constraint on
\(\Omega_{\rm tot}h^2\).
\textbf{Left:} Rescaled spin-independent cross section
\(\xi_{S_B}\sigma^{\rm SI}_{S_B}\) as a function of the WIMP mass. Grey
points fail the current LZ bound, while blue points satisfy the rescaled
direct-detection constraint and constitute the viable subset. The surviving
points cluster around the Higgs-resonance region,
\(M_{S_B}\simeq M_h/2\).
\textbf{Right:} Fractional contribution of the WIMP-like component,
\(\Omega_{S_B}/\Omega_{\rm tot}\). The remaining abundance can be supplied
by the feebly coupled \(S_A\) state with negligible direct-detection signal.}
    \label{fig:Z4FWS1}
\end{figure}

The relic-density-compatible parameter space and the impact of the
direct-detection constraint are shown in Fig.~\ref{fig:Z4FWS1}. The left panel
displays \(\xi_{S_B}\sigma^{\rm SI}_{S_B}\) as a function of the WIMP mass.
As in single-component Higgs-portal models, the viable subset is largely
concentrated around the Higgs-resonance region,
\(M_{S_B}\simeq M_h/2\)~\cite{Athron:2017Singlet}. In the present mixed setup, the
intra-dark interactions are restricted by the requirement that the \(S_A\)
population remain non-thermal, so the additional dark-sector channels do not
generically remove the usual Higgs-portal relic-density/direct-detection
correlation.

The right panel shows that \(S_B\) need not account for the full relic density.
Its fractional abundance,
\(\xi_{S_B}\equiv\Omega_{S_B}/\Omega_{\rm tot}\), is determined by the
coupled Boltzmann evolution, while freeze-in production of \(S_A\) can supply
the remaining abundance. Within the multi-component rescaling adopted in
Sec.~\ref{subsec:direct}, the direct-detection constraint therefore acts on
\(\xi_{S_B}\sigma^{\rm SI}_{S_B}\). A reduced WIMP fraction weakens the
effective direct-detection rate, but does not generically remove the usual
Higgs-portal correlation between freeze-out and scattering.

\begin{figure}[htbp]
    \centering
    \begin{subfigure}{1.0\textwidth}
        \centering
        \begin{minipage}{0.48\textwidth}
            \centering
            \includegraphics[width=\textwidth]{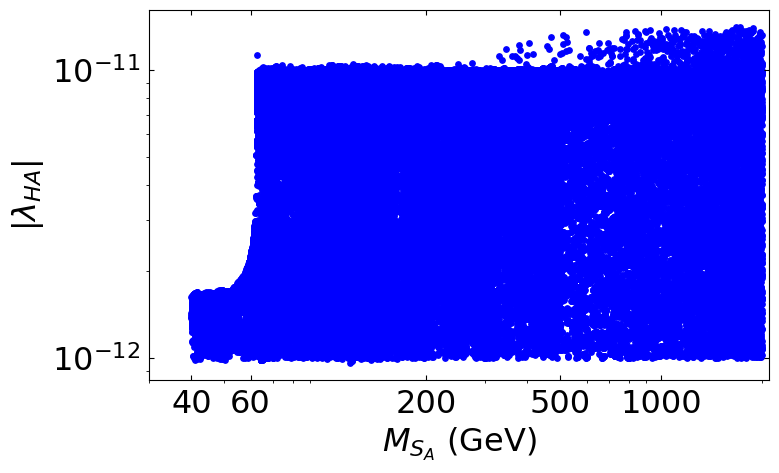}
        \end{minipage}
        \hfill
        \begin{minipage}{0.48\textwidth}
            \centering
            \includegraphics[width=\textwidth]{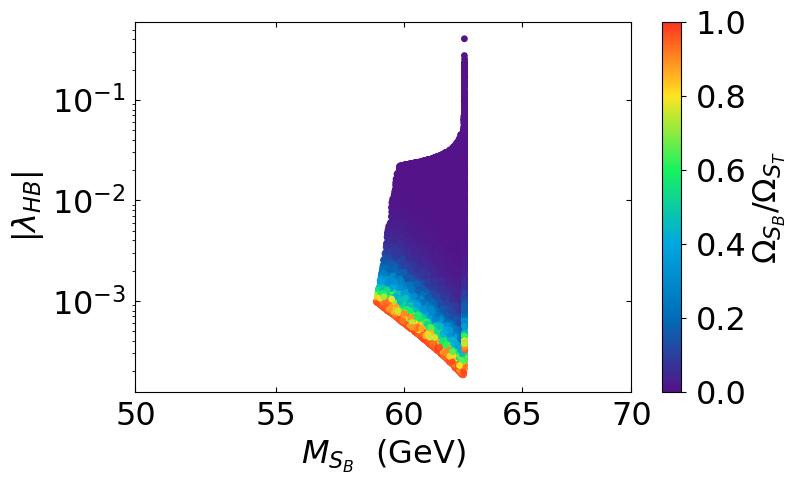}
        \end{minipage}
        \caption{\justifying Viable parameter space projected onto the
        Higgs-portal couplings for the mixed stable scenario with \(S_A\) as the
        FIMP-like component and \(S_B\) as the WIMP-like component. All points
        satisfy the Planck constraint on \(\Omega_{\rm tot}h^2\).
        \textbf{Left:} \(|\lambda_{HA}|\) versus \(M_{S_A}\). The freeze-in
        scaling is preserved, while the spread of viable points reflects that
        \(S_A\) is not required to saturate the relic density by itself.
        \textbf{Right:} \(|\lambda_{HB}|\) versus \(M_{S_B}\). Viable points
        concentrate near \(M_{S_B}\simeq M_h/2\), showing that the WIMP-like
        component remains controlled by the Higgs-funnel/direct-detection
        interplay.}
        \label{fig:Z4FWS2a}
    \end{subfigure}

    \vspace{1em}

    \begin{subfigure}{1.0\textwidth}
        \centering
        \begin{minipage}{0.48\textwidth}
            \centering
            \includegraphics[width=\textwidth]{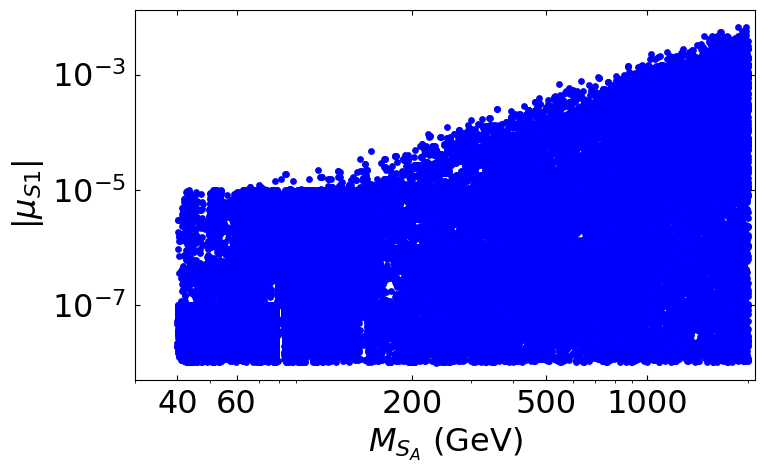}
        \end{minipage}
        \hfill
        \begin{minipage}{0.48\textwidth}
            \centering
            \includegraphics[width=\textwidth]{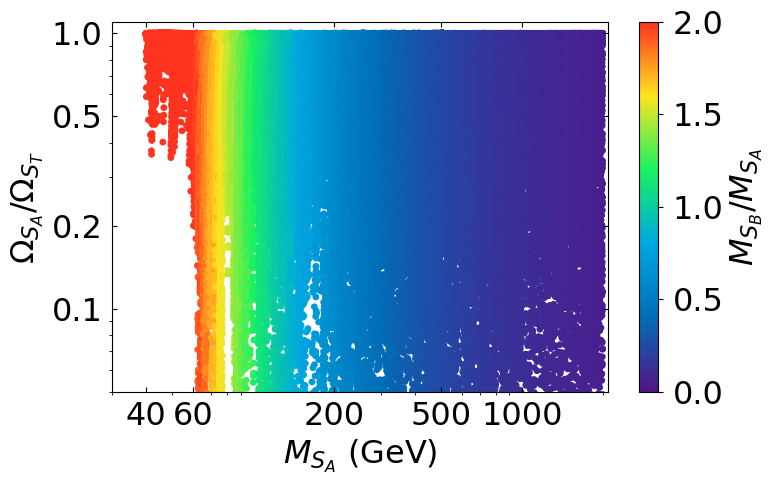}
        \end{minipage}
 \caption{\justifying Correlations involving the dark-sector interaction
and the relic-density composition in the mixed stable scenario.
\textbf{Left:} \(|\mu_{S1}|\) versus \(M_{S_A}\). The viable sample exhibits
a positive correlation between the FIMP mass and the trilinear coupling.
\textbf{Right:} \(\Omega_{S_A}/\Omega_{\rm tot}\) versus \(M_{S_A}\),
colour-coded by \(M_{S_B}/M_{S_A}\), showing that the FIMP component can
dominate over a broad mass range.}
        \label{fig:Z4FWS2b}
    \end{subfigure}
    \caption{\justifying Parameter-space correlations for the stable mixed
    scenario with \(S_A\) as a FIMP and \(S_B\) as a WIMP.}
    \label{fig:Z4FWS2}
\end{figure}

The portal-coupling projections are shown in Fig.~\ref{fig:Z4FWS2a}. The FIMP
portal magnitude \(|\lambda_{HA}|\) retains the characteristic freeze-in trend
familiar from single-component analyses~\cite{Yaguna:2011FIMP}, with additional
spread because \(S_A\) need not reproduce the full relic abundance by itself.
The WIMP portal \(|\lambda_{HB}|\), in contrast, retains the characteristic
Higgs-portal structure. The mixed scenario therefore allows the total relic abundance to be shared between the two states while leaving the direct-detection signal dominated by the WIMP-like component. This is expected because the FIMP-like state has a feeble Higgs-portal coupling, whereas the WIMP-like component retains the coupling required for thermal freeze-out. Its direct-detection contribution is evaluated using the relic-fraction rescaling adopted in Sec.~\ref{subsec:direct}.

Figure~\ref{fig:Z4FWS2b} shows additional correlations involving the
dark-sector interaction and the relic-density composition. The left panel exhibits a positive correlation between \(M_{S_A}\) and \(|\mu_{S1}|\) within the viable sample. This trend indicates that the relic-density and viability requirements select correlated values of the FIMP mass and the trilinear coupling, although it does not imply a one-to-one relation between them, since the final abundance depends on the coupled evolution and on the remaining model parameters. The right panel shows that \(S_A\) can dominate the total relic density across a broad mass range.

\subsubsection{Mixed stable scenario with a \texorpdfstring{\(S_A\)}{SA} WIMP component}
\label{subsubsec:wimpfimp}

We now consider the mixed stable configuration with the roles interchanged:
the complex scalar \(S_A\) is the thermal WIMP, coupled to the SM through
\(\lambda_{HA}\), while the real scalar \(S_B\) remains feebly interacting
and is produced through freeze-in via \(\lambda_{HB}\). The total relic
abundance can again be shared between the thermal and non-thermal components,
with the WIMP fraction
\(\xi_{S_A}\equiv\Omega_{S_A}/\Omega_{\rm tot}\) determined by the coupled
Boltzmann evolution rather than by requiring \(S_A\) alone to saturate the
Planck value. Within the relic-fraction rescaling adopted for direct
detection, the surviving WIMP-like points retain the characteristic
Higgs-portal correlation between freeze-out and spin-independent scattering
and are strongly concentrated near the Higgs resonance.

Within the surviving sample, this concentration has a direct kinematic
consequence. Stability requires \(M_{S_B}<2M_{S_A}\), and together with
\(M_{S_A}\simeq M_h/2\) this confines the FIMP-like \(S_B\) to
\(M_{S_B}\lesssim M_h\). The surviving points therefore place both dark
states in the electroweak-scale mass region, in contrast with the
\(S_B\)-WIMP/\(S_A\)-FIMP realisation, where the feebly coupled \(S_A\)
can extend to substantially larger masses. The FIMP-like \(S_B\) can
nevertheless provide the remaining relic abundance with a negligible
direct-detection contribution.

\begin{figure}[htbp]
    \centering
    \begin{subfigure}{0.48\textwidth}
        \centering
        \includegraphics[width=\textwidth]{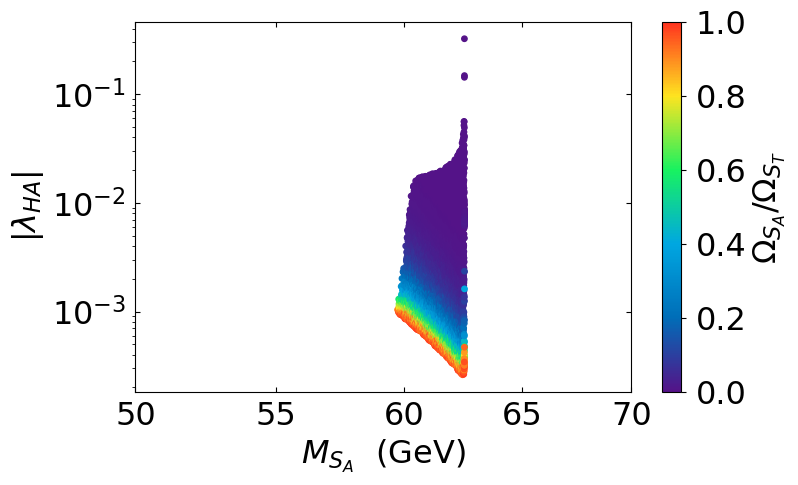}
    \end{subfigure}
    \hfill
    \begin{subfigure}{0.48\textwidth}
        \centering
        \includegraphics[width=\textwidth]{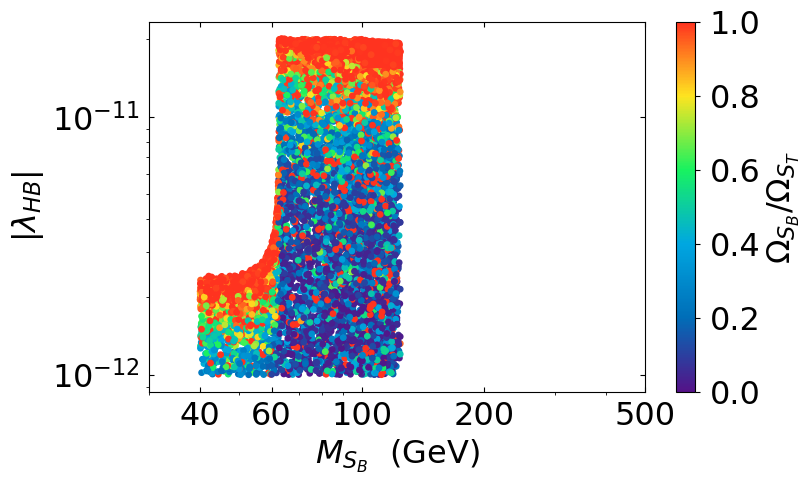}
    \end{subfigure}
    \caption{\justifying Viable parameter space for the mixed stable scenario
    with \(S_A\) as a WIMP and \(S_B\) as a FIMP.
    \textbf{Left:} \((M_{S_A},|\lambda_{HA}|)\) plane, colour-coded by the WIMP
    fractional abundance \(\Omega_{S_A}/\Omega_{\rm tot}\).
    \textbf{Right:} \((M_{S_B},|\lambda_{HB}|)\) plane, colour-coded by the FIMP
    fractional abundance \(\Omega_{S_B}/\Omega_{\rm tot}\).}
    \label{fig:Z4WFS}
\end{figure}

\subsection{Unstable Scenarios}
\label{subsec:unstablescenarios}

We now turn to the decay-mediated regime, defined by
\(M_{S_B}>2M_{S_A}\), for which the heavier state \(S_B\) is unstable and
\(S_B\to S_A S_A\) is kinematically open. The asymptotic dark matter
abundance is then carried by the stable \(S_A\) sector, while the decaying
\(S_B\) population can provide a non-thermal contribution whose importance
depends on the parent abundance at the decay epoch and on the corresponding
transfer of comoving number density into the stable daughter population.

To quantify the relevance of the decay for the final abundance of the stable
species, we use the diagnostic \(f_{\rm dec}\) defined in
Eq.~\eqref{eq:fdec}. For configurations exhibiting a distinct injection
epoch, it measures the relative increase of the \(S_A\) abundance between
the pre-injection plateau and the asymptotic post-injection value, and
therefore estimates the fraction of the final abundance supplied during
that epoch. We adopt \(f_{\rm dec}>0.1\) as an operational criterion to identify
configurations in which the decay-induced increase accounts for at
least 10\% of the final \(S_A\) yield. Points with
\(f_{\rm dec}<0.1\) are classified as having a subdominant contribution
according to this criterion. The threshold is used to organise and
display the numerical results and does not represent a universal
physical or observational boundary. Configurations in which \(S_B\) decays
before \(S_A\) freeze-out are treated separately, since the injected
population is subsequently processed by the thermal evolution and does not
generate a distinct late-injection contribution.

The thermal histories of the parent and daughter define three distinct
decay-mediated mechanisms. If \(S_B\) thermalises with the SM plasma,
freezes out, and subsequently decays into the feebly coupled \(S_A\), the
evolution realises the SuperWIMP mechanism. If instead \(S_B\) is produced
through freeze-in while \(S_A\) is thermal, its decay can inject an
additional population of \(S_A\) after freeze-out, leading to
injection-assisted freeze-out. Finally, when both states remain feebly
coupled, \(S_B\) is produced through freeze-in and subsequently feeds the
stable \(S_A\) population through its decay, resulting in sequential
freeze-in with an unstable parent.

\subsubsection{The SuperWIMP mechanism (FIMP--WIMP)}
\label{subsubsec:superwimp}

In the SuperWIMP regime~\cite{Feng:2003SuperWIMP_PRL,Feng:2003SuperWIMP_PRD}
(see also Refs.~\cite{Pukhov,YagunaZapata2024} for related mixed WIMP/FIMP
implementations), the parent state \(S_B\) lies in the WIMP-like portal range,
\(|\lambda_{HB}|\gtrsim10^{-4}\), and the accepted points are confirmed to
thermalise with the SM bath before undergoing freeze-out. Unlike a stable WIMP,
however, \(S_B\) subsequently decays into the feebly interacting daughter
\(S_A\), so the asymptotic dark matter abundance is entirely carried by the
stable \(S_A\) sector.

\begin{figure}[htbp]
    \centering
    \begin{subfigure}{0.48\textwidth}
        \centering
        \includegraphics[width=\textwidth]{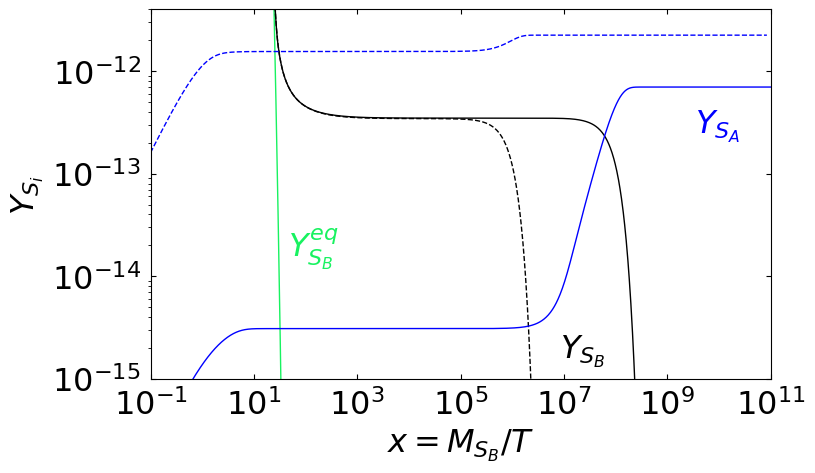}
    \end{subfigure}
    \hfill
    \begin{subfigure}{0.48\textwidth}
        \centering
        \includegraphics[width=\textwidth]{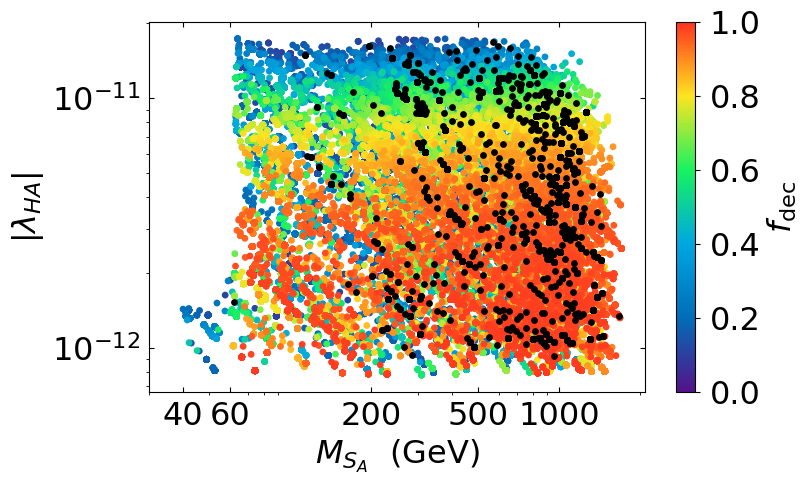}
    \end{subfigure}

    \vspace{0.5cm}
    \begin{subfigure}{0.58\textwidth}
        \centering
        \includegraphics[width=\textwidth]{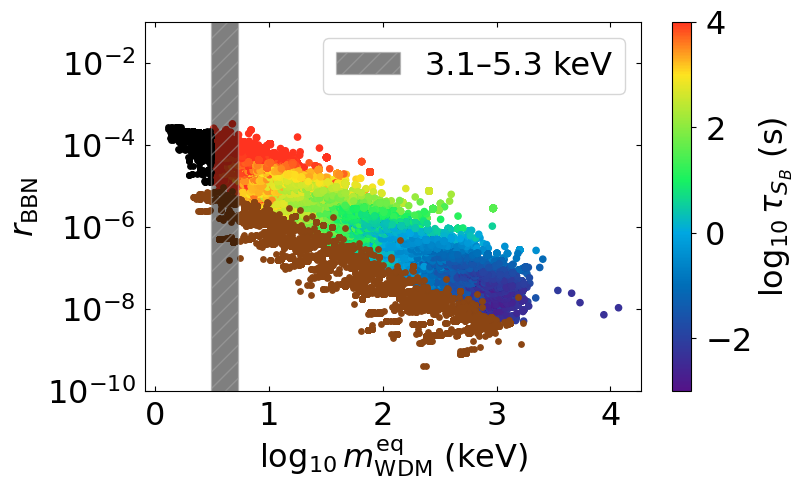}
    \end{subfigure}

\caption{\justifying
\textbf{Top left:} Evolution of comoving number densities in the SuperWIMP
scenario. The frozen-out abundance of the parent \(S_B\) (black) decreases
at late times due to decay, while the daughter abundance \(S_A\) (blue)
receives a corresponding non-thermal injection. The parameters are fixed to
\textit{Solid:} $M_{S_A} = 650.39$ GeV, $M_{S_B} = 1890.20$ GeV,
$\mu_{S1} = 4.98 \times 10^{-12}$ GeV,
$\lambda_{AB} = 1.03 \times 10^{-12}$,
$\lambda_{HA} = -1.01 \times 10^{-12}$, $\lambda_{HB} = 0.46$.
\textit{Dashed:} $M_{S_A} = 248.73$ GeV, $M_{S_B} = 850.18$ GeV,
$\mu_{S1} = 1.87 \times 10^{-10}$ GeV,
$\lambda_{AB} = -2.97 \times 10^{-11}$,
$\lambda_{HA} = -3.23 \times 10^{-12}$, $\lambda_{HB} = 0.32$.
\textbf{Top right:} Viable parameter space for the stable daughter \(S_A\).
\textbf{Bottom:} Cosmological diagnostic in the
\((r_{\rm BBN},m_{\rm WDM}^{\rm eq})\) plane, colour-coded by
\(\tau_{S_B}\). Black points satisfy \(f_{\rm dec}>0.1\) and
\(m_{\rm WDM}^{\rm eq}<3.1~\mathrm{keV}\), while brown points have
\(f_{\rm dec}<0.1\). The grey band,
\(3.1<m_{\rm WDM}^{\rm eq}<5.3~\mathrm{keV}\), indicates the range spanned
by the thermal-WDM reference limits used for comparison.}
    \label{fig:Z4FWUns}
\end{figure}

The final \(S_A\) abundance receives a direct freeze-in contribution through
the feeble portal \(\lambda_{HA}\) and an inherited contribution from the
frozen-out \(S_B\) population. The corresponding evolution is illustrated in
the top-left panel of Fig.~\ref{fig:Z4FWUns}: after freeze-out,
\(Y_{S_B}\) remains approximately constant until decays become efficient,
after which the parent abundance decreases while \(Y_{S_A}\) increases through
non-thermal injection. Assuming that the two CP-conjugate dark decay channels,
\(S_B\to S_A S_A\) and \(S_B\to S_A^\dagger S_A^\dagger\), saturate the
\(S_B\) decay width, each decay produces two stable dark-sector particles.
Neglecting post-injection depletion of \(S_A\), the resulting total relic
density of the stable \(S_A\) sector can be estimated as
\begin{equation}
\label{eq:superwimp_relic}
\Omega_{S_A} \simeq
\Omega^{\text{freeze-in}}_{S_A}
+ \frac{2 M_{S_A}}{M_{S_B}}\, \Omega^{\text{freeze-out}}_{S_B}.
\end{equation}
Here \(\Omega^{\text{freeze-out}}_{S_B}\) denotes the would-be relic density
associated with the frozen-out parent yield before decay. The second term
therefore represents the transfer of this frozen-out population into the
stable daughter sector.

The relic-density requirement and the direct-detection signal depend on
different aspects of the production history. The final \(\Omega_{S_A}\)
receives contributions from both direct freeze-in and the inherited parent
abundance, whereas direct detection probes the feeble Higgs portal of the
stable state, \(\lambda_{HA}\). As shown in the top-right panel of
Fig.~\ref{fig:Z4FWUns}, the contribution inherited from the WIMP-like
parent allows viable points at smaller \(|\lambda_{HA}|\) than would
typically be required if \(S_A\) had to account for the full abundance
through freeze-in alone~\cite{Yaguna:2011FIMP}. The corresponding
spin-independent scattering rates are consequently suppressed.

We now apply the cosmological diagnostics discussed in
Sec.~\ref{subsec:cosmoconstraints}. The bottom panel of
Fig.~\ref{fig:Z4FWUns} shows the distribution of viable points in the
\((r_{\rm BBN},m_{\rm WDM}^{\rm eq})\) plane, colour-coded by the parent
lifetime \(\tau_{S_B}\). We find \(r_{\rm BBN}\ll1\) throughout the retained sample, indicating
that the energy density of the metastable \(S_B\) population is
subdominant to radiation at the characteristic decay temperature
\(T_d\), within the adopted diagnostic. Since the decay proceeds
entirely within the dark sector, it does not directly inject
electromagnetic or hadronic energy into the visible plasma. The
warmness diagnostic, however, flags a significant fraction of the sample:
points shown in black satisfy \(f_{\rm dec}>0.1\) and
\(m_{\rm WDM}^{\rm eq}<3.1~\mathrm{keV}\), identifying configurations for
which a dedicated structure-formation analysis would be required.

A clear correlation is visible between \(\tau_{S_B}\) and
\(m_{\rm WDM}^{\rm eq}\). Longer-lived parents decay at lower temperatures,
leaving less time for the daughter momentum to redshift before the present
epoch. The resulting larger present-day velocity corresponds to a smaller
\(m_{\rm WDM}^{\rm eq}\), in accordance with
Eq.~\eqref{eq:mwdmeq}. The interval
\(3.1<m_{\rm WDM}^{\rm eq}<5.3~\mathrm{keV}\) corresponds to the spread of
thermal-WDM reference limits used for comparison. Equivalent masses within
this interval are therefore sensitive to the adopted thermal reference
rather than representing a model-independent exclusion.

\subsubsection{Injection-assisted freeze-out (WIMP--FIMP)}
\label{subsubsec:injection}

\begin{figure}[htbp]
    \centering
    \begin{subfigure}{0.48\textwidth}
        \centering
        \includegraphics[width=\textwidth]{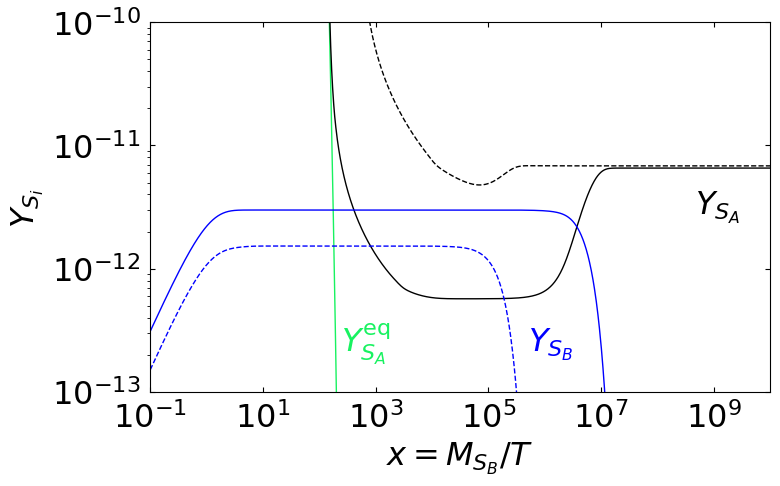}
    \end{subfigure}
    \hfill
    \begin{subfigure}{0.48\textwidth}
        \centering
        \includegraphics[width=\textwidth]{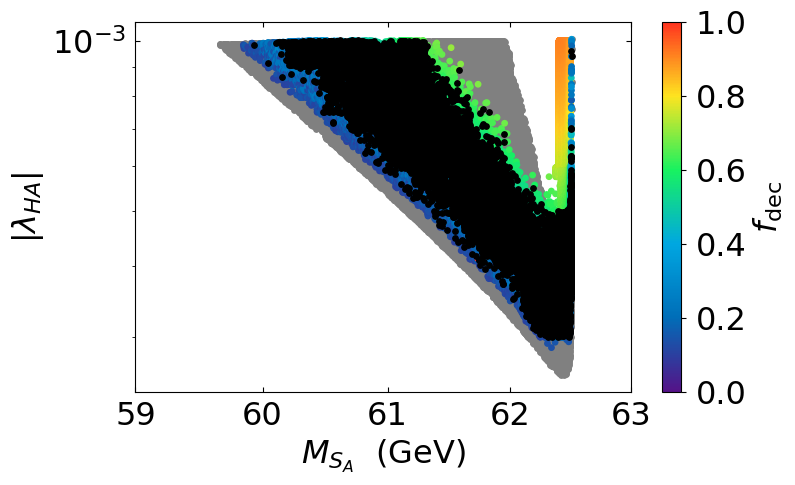}
    \end{subfigure}
\caption{\justifying Injection-assisted freeze-out scenario in which an
unstable FIMP parent \(S_B\) decays into a stable WIMP daughter \(S_A\).
\textbf{Left:} Evolution of comoving number densities. The WIMP-like
abundance of \(S_A\) (black) initially follows standard freeze-out and
subsequently receives a non-thermal contribution from the decay of the FIMP
parent \(S_B\) (blue). The parameters are fixed to
\textit{Solid:} $M_{S_A} = 62.45$ GeV, $M_{S_B} = 425.63$ GeV,
$\mu_{S1} = 7.17 \times 10^{-12}$ GeV,
$\lambda_{AB} = 1.43 \times 10^{-11}$,
$\lambda_{HA} = -9.95 \times 10^{-4}$,
$\lambda_{HB} = 3.63 \times 10^{-11}$.
\textit{Dashed:} $M_{S_A} = 62.49$ GeV, $M_{S_B} = 1597.5$ GeV,
$\mu_{S1} = 2.24 \times 10^{-9}$ GeV,
$\lambda_{AB} = 9.65 \times 10^{-12}$,
$\lambda_{HA} = 3.97 \times 10^{-4}$,
$\lambda_{HB} = 5.49 \times 10^{-11}$.
\textbf{Right:} Viable WIMP parameter space in the
\((M_{S_A},|\lambda_{HA}|)\) plane. Late-injection points are colour-coded
by \(f_{\rm dec}\). Black points satisfy \(f_{\rm dec}>0.1\) and
\(m_{\rm WDM}^{\rm eq}<3.1~\mathrm{keV}\). Grey points correspond to
configurations in which \(S_B\) decays before \(S_A\) freeze-out, so the
injected population is subsequently processed by the thermal evolution.}
    \label{fig:Z4WFUns}
\end{figure}

We next consider a feebly coupled parent \(S_B\) produced through freeze-in,
while the stable daughter \(S_A\) thermalises with the SM bath and undergoes
conventional freeze-out. If the decay \(S_B\to S_A S_A\) occurs sufficiently
late, the frozen-in parent population injects additional \(S_A\) particles
after the daughter has departed from thermal equilibrium, as illustrated in
the left panel of Fig.~\ref{fig:Z4WFUns}. The final \(S_A\) abundance then
depends both on its freeze-out dynamics and on the yield and lifetime of the
decaying FIMP parent.

Late injection modifies the relation between the Higgs-portal coupling
\(\lambda_{HA}\) and the final WIMP abundance. A parameter configuration
that would be underabundant under a single-component freeze-out
interpretation can become viable once the subsequent decay of the
frozen-in \(S_B\) population feeds additional \(S_A\) particles into the
post-freeze-out evolution. As shown in the right panel of
Fig.~\ref{fig:Z4WFUns}, the viable region remains closely associated
with the Higgs resonance in the \((M_{S_A},|\lambda_{HA}|)\) plane,
even though the final abundance is no longer determined by freeze-out
alone. The direct-detection signal remains controlled by the Higgs portal
of the WIMP-like daughter.

The right panel of Fig.~\ref{fig:Z4WFUns} also shows the impact of the
Lyman-\(\alpha\)-motivated warmness diagnostic. Late-injection points with
\(f_{\rm dec}>0.1\) and
\(m_{\rm WDM}^{\rm eq}<3.1~\mathrm{keV}\) are flagged as potentially warm,
while equivalent masses in the \(3.1\)--\(5.3~\mathrm{keV}\) interval remain
sensitive to the thermal-WDM benchmark adopted for comparison. Grey points
correspond instead to configurations in which \(S_B\) decays while \(S_A\)
is still in thermal equilibrium, so the injected population is processed by
the subsequent freeze-out evolution and does not survive as a distinct
non-thermal contribution. As in the SuperWIMP case, we find
\(r_{\rm BBN}\ll1\) throughout the viable sample, so the corresponding BBN
diagnostic is not shown separately.

\subsubsection{Sequential freeze-in (FIMP--FIMP)}
\label{subsubsec:sequential}

\begin{figure}[htbp]
    \centering
    \begin{subfigure}{0.48\textwidth}
        \centering
        \includegraphics[width=\textwidth]{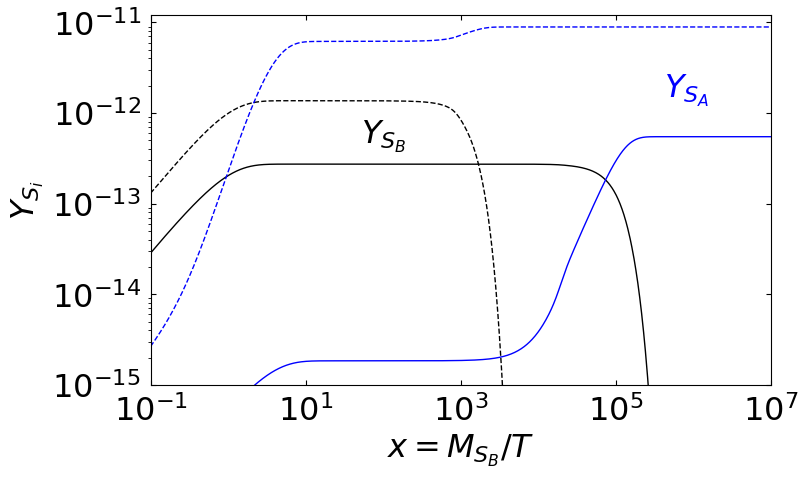}
    \end{subfigure}
    \hfill
    \begin{subfigure}{0.48\textwidth}
        \centering
        \includegraphics[width=\textwidth]{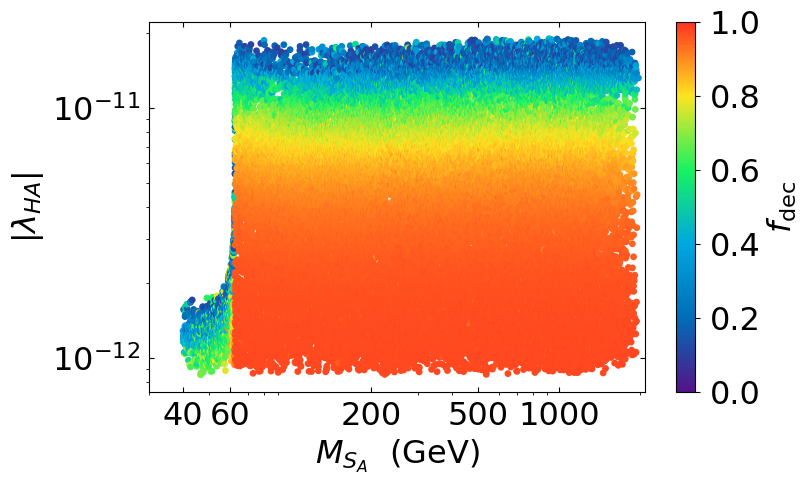}
    \end{subfigure}
 \caption{\justifying Sequential freeze-in scenario in which a feebly
produced parent \(S_B\) decays into the stable FIMP \(S_A\).
\textbf{Left:} Evolution of the \(S_A\) abundance. The parent \(S_B\) is
produced via freeze-in and later decays into \(S_A S_A\), yielding a delayed
non-thermal contribution in addition to the direct freeze-in production of
\(S_A\). The parameters are fixed to
\textit{Solid:} $M_{S_A} = 816.73$ GeV, $M_{S_B} = 2971.10$ GeV,
$\mu_{S1} = 1.06 \times 10^{-8}$ GeV,
$\lambda_{AB} = 2.39 \times 10^{-9}$,
$\lambda_{HA} = 1.02 \times 10^{-12}$,
$\lambda_{HB} = 3.37 \times 10^{-11}$.
\textit{Dashed:} $M_{S_A} = 53.01$ GeV, $M_{S_B} = 137.27$ GeV,
$\mu_{S1} = 1.10 \times 10^{-8}$ GeV,
$\lambda_{AB} = -3.89 \times 10^{-11}$,
$\lambda_{HA} = -1.45 \times 10^{-12}$,
$\lambda_{HB} = 1.52 \times 10^{-11}$.
\textbf{Right:} Viable parameter space for the stable FIMP \(S_A\),
colour-coded by \(f_{\rm dec}\).}
    \label{fig:Z4FFUns}
\end{figure}

In this realisation both dark states remain out of equilibrium with the SM
bath. The stable FIMP \(S_A\) receives a direct freeze-in contribution through
\(\lambda_{HA}\) and a sequential contribution from the parent \(S_B\), which
is itself produced via freeze-in and subsequently decays into \(S_A S_A\).
The resulting two-step evolution is illustrated in the left panel of
Fig.~\ref{fig:Z4FFUns}.

The parent contribution modifies the relation between the FIMP mass and
its Higgs-portal coupling that would arise if \(S_A\) were produced solely
through direct freeze-in. The final \(S_A\) abundance receives contributions
from both direct freeze-in and the decay of the freeze-in-produced \(S_B\)
population. For sufficiently large \(f_{\rm dec}\), the inherited component
provides a sizeable fraction of the final abundance. The corresponding
viable parameter space is shown in the right panel of
Fig.~\ref{fig:Z4FFUns}, colour-coded by \(f_{\rm dec}\). The daughter
momentum at production is fixed by the decay kinematics, while the decay
epoch controls its subsequent cosmological redshift and therefore enters
the warmness diagnostic.

Unlike in the previous two decay-mediated scenarios, the adopted
warmness diagnostic does not flag any points in the FIMP--FIMP regime.
Even for configurations with large \(f_{\rm dec}\), we find
\(m_{\rm WDM}^{\rm eq}\gtrsim1~\mathrm{MeV}\), well above the thermal-WDM
reference range used for comparison. This conclusion applies to the
velocity-matching proxy of Sec.~\ref{subsec:cosmoconstraints} and does
not replace a full transfer-function calculation or a Lyman-\(\alpha\)
likelihood analysis. Within this diagnostic, the result reflects the
combined effect of the daughter momentum at production and its
subsequent cosmological redshift, encoded respectively through
\(p_\star\) and \(T_d\), rather than the smallness of the portal
couplings alone. We also find \(r_{\rm BBN}\ll1\) for all retained
points.

\section{Conclusions}
\label{sec:conclusions}

Our analysis shows that the viable parameter space and relic-density
composition of the minimal two-component \(\mathbb{Z}_4\) Higgs-portal
model depend not only on the masses and interaction strengths, but also
on the cosmological production history of the dark states. Allowing either
or both states to remain out of equilibrium opens relic-density histories
beyond the two-WIMP case without generically removing the usual
phenomenological constraints. Instead, non-thermal production changes how
the observed abundance is distributed between the two dark states and, when
\(S_B\) decays, how that abundance is transferred.

This distinction is already evident when both particles are stable.
When both states are FIMP-like, their abundances are generated
predominantly through their respective freeze-in channels. The requirement
that their combined abundance reproduce the observed relic density
selects correlated combinations of masses and portal couplings.
In the mixed WIMP--FIMP regimes, the feebly coupled component can provide a
substantial part of the abundance with a negligible direct-detection rate.
The WIMP-like component nevertheless retains the usual Higgs-portal
connection between freeze-out and spin-independent scattering, and the
surviving parameter space remains strongly associated with the Higgs
resonance. Interchanging which scalar is thermal also changes the viable mass
spectrum. When \(S_A\) is WIMP-like, its concentration near the Higgs
resonance, together with the stability condition
\(M_{S_B}<2M_{S_A}\), restricts the viable \(S_B\) masses to below
approximately \(M_h\) in the scanned sample. When \(S_B\) is
WIMP-like, the FIMP-like \(S_A\) can extend to substantially larger
masses, up to the upper end of the scan.

For \(M_{S_B}>2M_{S_A}\), the heavier scalar can decay into \(S_A\),
and its contribution to the final abundance depends on how both states
were produced. A parent that freezes out before decaying produces
\(S_A\) through the SuperWIMP mechanism. If \(S_B\) is instead produced
through freeze-in, its decay can increase the abundance of a WIMP-like
\(S_A\) after freeze-out, or contribute to sequential freeze-in when
\(S_A\) is also FIMP-like. When the stable daughter is FIMP-like and
\(f_{\rm dec}\) is sizeable, part of its final abundance comes from
the decaying parent. As a result, less \(S_A\) needs to be produced
through its own Higgs portal than in a scenario dominated by direct
freeze-in.

Late dark-sector decays introduce a complementary cosmological diagnostic.
Since the primary \(S_B\) decay products remain in the dark sector, the
decay does not directly inject electromagnetic or hadronic energy into the
visible plasma. Independently, we find \(r_{\rm BBN}\ll1\) throughout the
retained sample, indicating that the parent energy density is subdominant
to radiation at the characteristic decay temperature \(T_d\), according to
the diagnostic defined in Eq.~\eqref{eq:rBBN}.

For structure formation, we use \(f_{\rm dec}\) together with the
equivalent thermal mass \(m_{\rm WDM}^{\rm eq}\) as an approximate
velocity-matching diagnostic for the non-thermal daughter population.
Configurations with \(f_{\rm dec}>0.1\) and
\(m_{\rm WDM}^{\rm eq}<3.1~\mathrm{keV}\) are flagged as potentially warm,
while the interval \(3.1\)--\(5.3~\mathrm{keV}\) reflects the spread of
representative thermal-WDM reference limits rather than a
model-independent exclusion. In the sequential freeze-in sample, we instead
find \(m_{\rm WDM}^{\rm eq}\gtrsim1~\mathrm{MeV}\), so no points receive
this warmness flag. A quantitative Lyman-\(\alpha\) constraint on the
decay-produced population would require its full non-thermal phase-space
distribution and the corresponding matter transfer function.

By comparing these six production histories within the same model,
we can identify how the relic abundance and the viable parameter space
change when the assumption of thermal equilibrium is relaxed. The results
show that Higgs-portal constraints remain important for thermal components,
while dark-sector decays can change the final abundance and introduce
additional questions for structure formation. Both effects must be taken
into account when assessing the phenomenology of this
\(\mathbb{Z}_4\) model.

\section*{Acknowledgments}

We thank A.~Pukhov for useful discussions and for assistance with \texttt{micrOMEGAs}.
B.~L.~S\'anchez-Vega acknowledges financial support from the Minas Gerais State Research
Support Foundation, FAPEMIG, under Grant No.~APQ-02289-26. J.~P.~Carvalho-Corr\^ea acknowledges financial support from the National Council for Scientific
and Technological Development (CNPq) through Grant No.~141118/2022-9.
B.~A.~Couto e Silva acknowledges financial support from the Funda\c{c}\~ao de Amparo \`a Pesquisa do Estado de Minas
Gerais (FAPEMIG) and from the Coordena\c{c}\~ao de Aperfei\c{c}oamento de Pessoal de N\'ivel Superior (CAPES).

\appendix

\section{Theoretical Constraints}
\label{app:unitarity}

In this appendix we summarise the bounded-from-below (BFB), tree-level
perturbative-unitarity, and perturbativity conditions imposed in the numerical
analysis. The derivation of the BFB conditions and the corresponding
high-energy scalar-scattering constraints for the \(\mathbb{Z}_4\) framework
can be found in Ref.~\cite{CarvalhoCorrea:2025Z2n} and references therein.
The dimensionful trilinear parameter \(\mu_{S1}\) does not contribute to the
asymptotic BFB conditions or to the high-energy scalar-scattering eigenvalues
considered below.

\subsection{Perturbative unitarity}
\label{app:partialwave}

Tree-level perturbative unitarity is imposed by requiring the zeroth partial-wave
amplitude for all \(2\to2\) scalar scattering processes to satisfy
\(|\Re(a_0)|\leq 1/2\). Following the conventions of
Ref.~\cite{CarvalhoCorrea:2025Z2n}, this condition is implemented as
\begin{equation}
\label{eq:unitarity_general_bound_app}
|\mathcal{M}_i|\leq 8\pi,
\end{equation}
for each eigenvalue \(\mathcal{M}_i\) of the high-energy scalar-scattering
matrices.

For the scalar potential in Eq.~\eqref{eq:pot}, this gives
\begin{equation}
\label{eq:unitarity_bounds_simple}
|\lambda_H|,\,|\lambda_{AB}|,\,|\lambda_A \pm 3\lambda_{S4}|\leq 4\pi,
\qquad
|\lambda_{HA}|,\,|\lambda_{HB}|\leq 8\pi.
\end{equation}

The coupled neutral sector spanned by
\((hh,\;S_A S_A^\dagger,\;S_B S_B)\) yields three eigenvalues
\(x_{1,2,3}\), determined by the roots of
\begin{equation}
\label{eq:unitarity_cubic}
x^3 + A\,x^2 + B\,x + C = 0,
\end{equation}
with
\begin{align}
A &= 12\lambda_H + 8\lambda_A + 24\lambda_B,
\label{eq:unitarity_coeffA}
\\
B &= 96\lambda_H\lambda_A
+ 288\lambda_H\lambda_B
+ 192\lambda_A\lambda_B
- 8\lambda_{AB}^2
- 8\lambda_{HA}^2
- 4\lambda_{HB}^2,
\label{eq:unitarity_coeffB}
\\
C &= 2304\lambda_H\lambda_A\lambda_B
- 96\lambda_H\lambda_{AB}^2
- 32\lambda_A\lambda_{HB}^2
- 192\lambda_B\lambda_{HA}^2
+ 32\lambda_{AB}\lambda_{HA}\lambda_{HB}.
\label{eq:unitarity_coeffC}
\end{align}

We require
\begin{equation}
\label{eq:unitarity_roots_bound}
|x_k|\leq 16\pi,
\qquad
k=1,2,3,
\end{equation}
using the same normalisation and identical-particle conventions as in
Ref.~\cite{CarvalhoCorrea:2025Z2n}.

\subsection{Global bounded-from-below condition}
\label{app:bfb_global}

In addition to the field-axis and pairwise BFB conditions in
Eqs.~\eqref{eq:estabilidade_diagonal}--\eqref{eq:condHB}, one must require the
quartic potential to remain non-negative when the three field directions
\(H\), \(S_A\), and \(S_B\) are simultaneously nonzero. Using the effective
couplings \(\overline{\lambda}_{ij}\) defined in
Eqs.~\eqref{eq:condAB}--\eqref{eq:condHB}, the corresponding copositivity
condition is
\begin{equation}
\label{eq:BFB_global}
\begin{split}
2\lambda_{AB}\sqrt{\lambda_{H}}
&+ 2\lambda_{HA}\sqrt{\lambda_{B}}
+ \lambda_{HB}\sqrt{\lambda_{A}-|\lambda_{S4}|} \\
&+ 4\sqrt{\lambda_{H}(\lambda_{A}-|\lambda_{S4}|)\lambda_{B}}
+ \sqrt{2\,\overline{\lambda}_{HA}\,
           \overline{\lambda}_{HB}\,
           \overline{\lambda}_{AB}}
>0 .
\end{split}
\end{equation}

Together with the field-axis and pairwise BFB conditions quoted in
Sec.~\ref{sec:model}, Eq.~\eqref{eq:BFB_global} implements the strict
copositivity requirement for the quartic potential and ensures positivity
along arbitrary large-field directions.

\subsection{Perturbativity}
\label{app:perturbativity}

Finally, we impose the perturbativity cut
\begin{equation}
\label{eq:perturbativity}
|\lambda_i|\leq 4\pi,
\qquad
\lambda_i\in
\{\lambda_H,\lambda_A,\lambda_B,\lambda_{HA},\lambda_{HB},
\lambda_{AB},\lambda_{S4}\}.
\end{equation}

All points failing any of the theoretical constraints listed in this appendix
are discarded from the scan.

\section{A posteriori thermalisation check}
\label{subsec:thermalisation_check}

The WIMP/FIMP labels used in the numerical scan are initially assigned through
the coupling ranges in Table~\ref{tab:scan_params}. To verify that these labels
correspond to the actual cosmological evolution, we perform an a posteriori
thermalisation check based on the abundance-changing interaction rates relevant
for chemical equilibration.

For each component classified as FIMP-like, denoted generically by \(F\), we
define the total abundance-changing interaction rate
\begin{equation}
\Gamma_F^{\rm tot}(x_F)
\equiv
\sum_{c\in\mathcal C_F}\Gamma_F^{(c)}(x_F),
\qquad
x_F\equiv\frac{M_F}{T},
\end{equation}

where \(\mathcal C_F\) denotes the set of reaction classes capable of changing
the abundance of \(F\), and \(\Gamma_F^{(c)}\) is the corresponding thermally
averaged interaction rate.

We then define the maximal rate-to-expansion ratio
over the temperature interval used in the numerical check,
\begin{equation}
R_F^{\rm max}
\equiv
\max_{x_F\in[10^{-2},\,25]}
\left[
\frac{\Gamma_F^{\rm tot}(x_F)}{H(x_F)}
\right].
\end{equation}

The rates are evaluated numerically using the channels generated by the model
implementation and include Higgs-portal production, dark-sector conversion,
semi-annihilation, and inverse decays when kinematically relevant. The interval
\(x_F\in[10^{-2},25]\) extends from the relativistic regime through the
epochs relevant for freeze-in production and thermal freeze-out.

A point is retained in the FIMP-like sample only if
\begin{equation}
R_F^{\rm max}<1.
\end{equation}

Thus, throughout the temperature interval covered by the diagnostic, the total
abundance-changing interaction rate of the nominal FIMP remains below the
Hubble expansion rate. We use this as the operational criterion that the
species does not reach chemical equilibrium with the SM bath or with a
thermalised dark-sector component.

Conversely, for a component classified as WIMP-like, denoted generically by
\(W\), we verify that abundance-changing interactions establish thermal contact
before freeze-out. Using the corresponding total interaction rate
\(\Gamma_W\), we require
\begin{equation}
\frac{\Gamma_W}{H}>1
\qquad
\text{before freeze-out}.
\end{equation}

In practice, the retained WIMP-like points satisfy
\(\Gamma_W/H\gg1\) during the early thermal evolution, so their classification
does not rely on configurations marginally close to the equilibration
threshold.

In the decay-mediated scenarios with \(M_{S_B}>2M_{S_A}\), the criterion is
applied with the appropriate time ordering. Inverse decays and scatterings
capable of establishing chemical equilibrium at early times are included in
the thermalisation diagnostic. By contrast, a late one-way injection from an
already frozen-out or feebly produced parent is not itself interpreted as
thermalisation of the daughter once the inverse processes are inefficient.
Such decays constitute the intended SuperWIMP or sequential-freeze-in
production mechanism. The check therefore rejects configurations in which a
nominally FIMP-like state is driven into equilibrium while retaining
genuinely non-thermal decay-mediated histories.

\bibliographystyle{utphys}
\bibliography{references}

\end{document}